\documentclass[aps,epsfig,multicol,amstex]{revtex4}
\usepackage{graphicx}
 \usepackage{amsmath}
 \usepackage{color}
\usepackage[linesnumbered]{algorithm2e} 
\usepackage{lmodern,amsmath,amssymb}
\usepackage{rotating}
\usepackage{todo}

\newcommand{\E}[1]{\mathbb E\left[#1\right]}
\newcommand{\bo}[1]{\boldsymbol{#1}}

\usepackage{ulem}
\usepackage{xcolor}
\newcommand\wolf[1]{{ #1}}
 
\newcommand{\CR}[1]{{\color{black}{#1}}}

\begin{document}
\title{
Parsimonious Modelling with  Information Filtering Networks
}

\author{
Wolfram Barfuss$^{1*}$,
Guido Previde Massara$^2$,
T. Di Matteo$^{2,3,4}$ and
Tomaso Aste$^{2,4}$}
\affiliation{
$^1$ Department of Physics, FAU Erlangen-N\"urnberg, N\"{a}gelsbachstr. 49b, 91052 Erlangen, GER \\
$^2$ Department of Computer Science, University College London,  Gower Street, London, WC1E 6BT, UK\\ 
$^3$ Department of Mathematics, King's College London,  The Strand, London, WC2R 2LS, UK\\
$^4$ Systemic Risk Centre, London School of Economics and Political Sciences, London, WC2A2AE, UK\\
$^*$now at: Potsdam Institute for Climate Impact Research, Telegrafenberg A31, 14473 Potsdam, GER\\
and Department of Physics, Humboldt University, Newtonstr. 15, 12489 Berlin, GER}
\date{\today}

\begin{abstract}
We introduce a methodology to construct \CR{parsimonious probabilistic} models. 
This method \CR{makes use of   Information Filtering Networks to produce a robust estimate of } the global sparse inverse covariance from a simple sum of local inverse covariances computed on small sub-parts of the network.
Being based on  local and low-dimensional inversions,  this method is computationally very efficient and statistically robust even for the estimation of inverse covariance of high-dimensional, noisy and short time-series.
\CR{Applied to financial data our method results computationally more efficient  than state-of-the-art methodologies such as Glasso producing, in a fraction of the computation time, models that can have equivalent or better performances but with a sparser  inference structure.}
\CR{We also discuss performances  with sparse factor models where we notice that relative performances decrease with the number of factors.}
The local nature of this approach allows us to perform computations in parallel and provides a tool for dynamical adaptation by partial updating when the properties of some variables change without the need of recomputing the whole model.
This makes this approach particularly suitable to handle big datasets with large numbers of variables.
\CR{Examples of practical application for forecasting, stress testing and risk allocation in financial systems are also provided.}
\end{abstract}

\maketitle

\section{Introduction}

This paper addresses the following question: how can one construct, from a set of observations,  the model that most meaningfully describes the underlying system? 
\CR{This is a general question at the core of scientific research. 
Indeed, the so-called scientific method has been devised around a combination of observation, model and prediction in a circular way, where hypotheses are formulated and tested with further observations, iteratively refining or changing the models to obtain better predictions; all within the principle of parsimony where a simpler model with less parameters and less assumptions should be preferred to a more complex one.
}

In the context of the present paper the `model' is the multivariate probability distribution that best describes the set of observations.
The problem of finding such a distribution becomes particularly challenging when the number of variables, $p$, is large and the number of observations, $q$, is small.
Indeed, in such a multivariate problem, the model must take into account at least an order $\mathcal{O}(p^2)$ of interrelations between the variables and therefore the number of model-parameters scales at least quadratically with the number of variables. 
A parsimonious approach requires to discover the model that best reproduces the statistical properties of the observations while keeping the number of parameters as small as possible.
Using a maximum entropy approach, up to  the second order in the moments of the distribution, the model becomes the multivariate normal distribution. In the multivariate normal case there is a simple relationship between the sparsity pattern of the inverse of the covariance matrix (the precision matrix, henceforth denoted by $\mathbf{J}$) and the underlying partial correlation structure (referred to as `graphical model' in the literature \cite{lauritzen1996}): two nodes $i$ and $j$ are linked in the graphical model if and only if the corresponding precision matrix element ${J}_{ij}$ is different from zero.
Therefore the problem of estimating a sparse precision matrix is equivalent to the problem of learning a sparse multivariate normal graphical model (known in the literature as Gaussian Markov Random Field (GMRF) \cite{rueheld2005}). Once the sparse precision matrix has been estimated, a number of efficient tools -- mostly based on research in sparse numerical  linear algebra  -- can be used to sample from the distribution, calculate conditional probabilities, calculate conditional statistics and forecast \cite{davis2006direct,rueheld2005}.
GMRFs are of great importance in many applications spanning computer vision \cite{Wang:2013:MRF:2527819.2528182}, sparse sensing \cite{montanari2012graphical}, finance  \cite{tumminelloetal2005,NJP10,pozzi2013spread,musmeci2015relation,musmeci2015risk,denevpgm}, gene expression 
\cite{WuEtAl2003,schaeferstrimmer2005,LezonEtAl2006}; biological neural networks \cite{SchneidmanEtAl2006}, climate networks \cite{ZerennerEtAl2014,EbertUphoffDeng2012_A}; geostatistics and spatial statistics \cite{haran2011gaussian, HristopulosElogne2009,Hristopulos2003}. Almost universally, applications require modelling a large number of variables with a relatively small number of observations and therefore the issue of the statistical significance of the model parameters is very important. 

The problem of finding meaningful and parsimonious models, sometimes referred as sparse structure learning \cite{zhou2011structure}, has been tackled by using a number of different approaches.
Let us hereafter briefly account for some of the most relevant in the present context.

\textit{Constraint based approaches} recover the structure of the network by testing the local Markov property. 
Usually the algorithm starts from a complete model and adopts a \textit{backward selection} approach by testing the independence of nodes conditioned on subsets of the remaining nodes (algorithms SGS and PC \cite{zhou2011structure}) and removing edges associated to nodes that are conditionally independent; the algorithm stops when some criteria are met -- e.g. every node has less than a given number of neighbours. Conversely \textit{forward selection} algorithms start from a sparse model and add edges associated to nodes that are discovered to be conditionally dependent. 
An hybrid model is the GS algorithm where  candidate edges are added to the model (the ``grow'' step) in a forward selection phase and subsequently reduced using a backward selection step (the ``shrinkage'' step) \cite{zhou2011structure}.
However, the complexity of checking a large number of conditional independence statements makes these methods unsuitable for moderately large graphs.
Furthermore, aside from the complexity of measuring conditional independence, these methods do not generally optimize a global function, such as likelihood or the Akaike Information Criterion \cite{akaike1974new,akaike1998information} but they rather try to exhaustively test all the (local) conditional independence properties of a set of data and therefore are difficult to use in a probabilistic framework. 

\textit{Score based approaches} learn the inference structure trying to optimize some global function: likelihood, Kullback-Leibler divergence \cite{kullback1951information}, Bayesian Information Criterion (BIC) \cite{schwarz1978estimating}, Minimum Description Length \cite{rissanen1978modeling} or the likelihood ratio test statistics \cite{petitjean2013scaling}. 
In these approaches, the main issue is that the optimization is generally computationally demanding and some sort of greedy approach is required. 
\CR{Statistical physics methodologies for the discovery of inference networks by maximization of entropy or minimization of cost functions have been used in biologically motivated studies to extract  gene regulatory networks  and signalling networks  (see for instance \cite{Molinelli13,Diambra11}). }

In the field of \textit{decomposable models} there are a number of methods that efficiently explore the graphical structure (directed, in the case of Bayesian models, or undirected in the case of log-linear or multivariate Gaussian models) by using advanced graph structures such as junction tree or clique graph  \cite{giudici1999decomposable,deshpande2001efficient,petitjean2013scaling}, with the goal of producing sparse models (so-called ``thin junction trees''  \cite{bach2001thin,srebro2001maximum,chechetka2008efficient}).

Other approaches \cite{d2008first, banerjee2008model,banerjee2006convex} treat the problem as a constrained optimization problem to recover the sparse covariance matrix. 
Within this line,  \textit{regression based approaches} generally try to minimize some loss function which enforces parsimony and sparsity by using penalization to constrain the number and size of the regression parameters. 
Specifically ridge regression uses a $\ell_2$-norm penalty; instead the \textit{lasso} method  \cite{Tibshirani1996}  uses an $\ell_1$-norm penalty and the elastic-net approach uses a convex combination of $\ell_2$ and $\ell_1$ penalties on the regression coefficients \cite{zou2005regularization}. These approaches are among the best performing regularization methodologies presently available. The $\ell_1$-norm penalty term favours solutions with parameters with zero value leading to models with sparse inverse covariances. Sparsity is controlled by regularization parameters $\lambda_{ij} > 0$; the larger the value of the parameters the more sparse the solution becomes.
This approach has become extremely popular and, around the original idea, a large body of literature has been published with several novel algorithmic techniques that are continuously advancing this method \cite{Tibshirani1996,MeinshausenBuehlmann2006,BanerjeeEtAl2008,RavikumarEtAl2011,HsiehEtAl2011,OztoprakEtAl2012}  among these the popular implementation \emph{Glasso} (Graphical-lasso)  \cite{FriedmanEtAl2008} which uses lasso to compute sparse graphical models.
However, Glasso methods are computationally intensive and, although they are sparse, the non-zero parameters interaction structure tends to be noisy and not significantly related with the true underlying interactions between the variables.

\begin{figure}[t]
\centering
\includegraphics[width=0.7\textwidth]{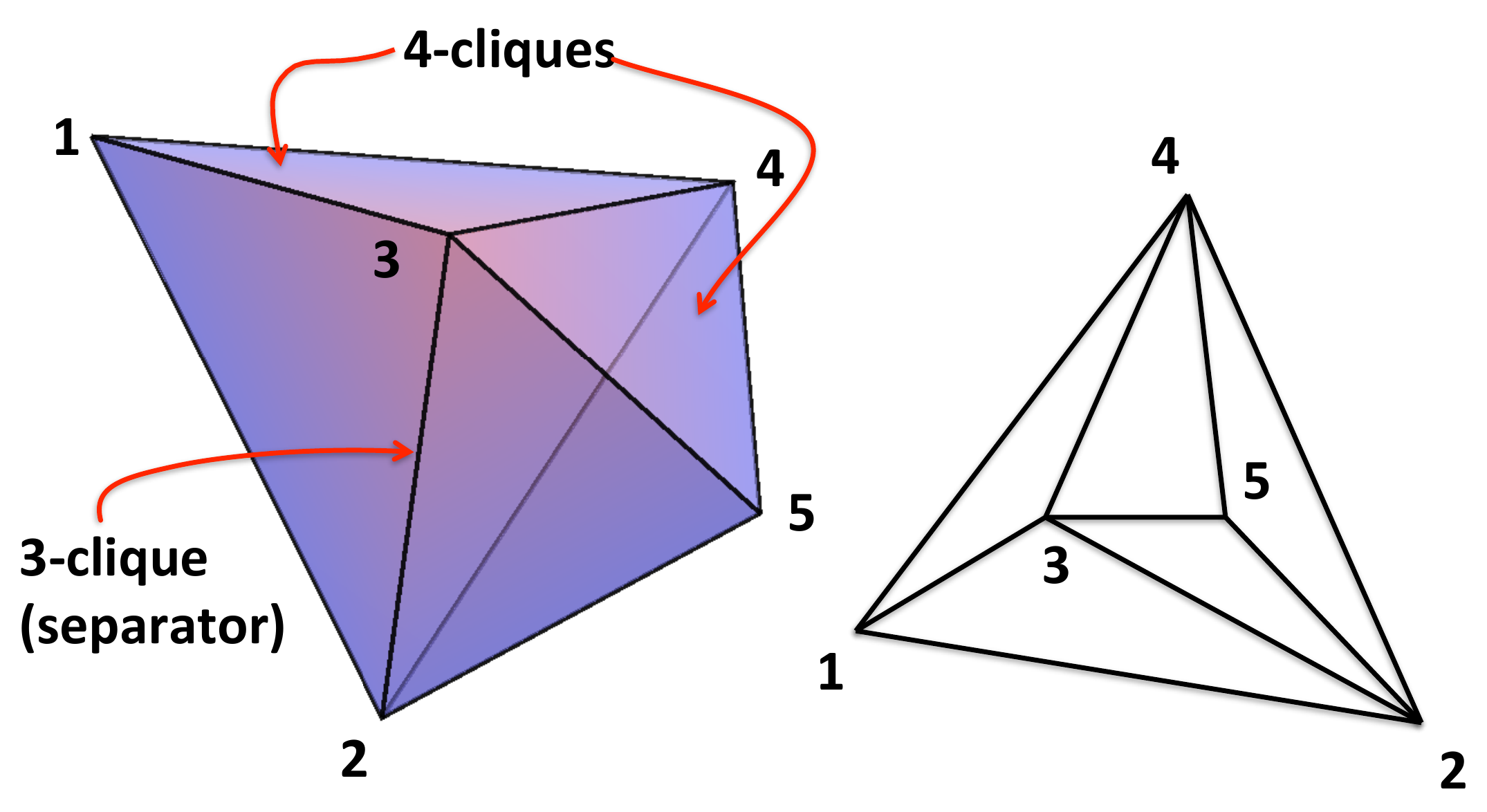}
\caption{
{\bf Two schematic representations of a TMFG network}. The network is made of two cliques of four nodes (4-cliques) and one separator of three nodes (3-clique). The cliques are the tetrahedra $\{1,2,3,4\}$ and $\{2,3,4,5\}$  and the separator is the triangle  $\{2,3,4\}$.
This is a chordal graph. 
}
\label{Fig:cliquesAndSpearator}
\end{figure}

\CR{Meaningful structures associated with the relevant network of interactions in complex systems are instead retrieved by   Information Filtering Networks which were first introduced by Mantegna \cite{mantegna1999} and some of the authors of the present paper \cite{tumminelloetal2005,AsteNetw06} for the study of the structure of financial markets and biological systems \cite{song2012hierarchical}. There is now a large body of literature  demonstrating  that   Information Filtering Networks such as Maximum Spanning Tree (MST) or Planar Maximally Filtered Graphs (PMFG) constructed from correlation matrices retrieve meaningful structures, extracting the relevant interactions from multivariate datasets in complex systems \cite{NJP10,song2012hierarchical,pozzi2013spread}.}
Recently, a new family of   Information Filtering Networks, the triangulated maximal planar graph (TMFG) \cite{TMFG}, was  introduced.  These are planar graphs, similar to the PMFG, but with the advantages to be generated in a computationally efficient way and, more importantly for this paper, they are decomposable graphs (see example in Fig.\ref{Fig:cliquesAndSpearator}). 
A decomposable graph has the property that every cycle of  length greater than three has a chord, an edge that connects two vertices of the cycle in a smaller cycle of length three. 
Decomposable graphs, also called chordal or triangulated, are clique forests, made of k-cliques (complete sub graphs of k vertices) connected by separators.
Separators are also cliques of smaller sizes with the property that the graph becomes divided into two or more disconnected components when the vertices of the separator are disconnected.
For example, in the schematic representation of the TMFG reported in Fig.\ref{Fig:cliquesAndSpearator} the cliques are  the tetrahedra $\{1,2,3,4\}$ and $\{2,3,4,5\}$ whereas the separator is the triangle  $\{2,3,4\}$.

The novelty of the method presented in this paper is the combination of decomposable  Information Filtering Networks
\cite{tumminelloetal2005,NJP10,pozzi2013spread,musmeci2015relation,musmeci2015risk} with Gaussian Markov Random Fields \cite{rueheld2005,lauritzen1996} to produce parsimonious models associated with a meaningful structure of dependency between the variables. 
The strength of this methodology is that the global sparse inverse covariance matrix is produced from a simple sum of local inversions.
This makes the method computationally very efficient and statistically  robust.
Given the Local/Global nature of its construction, in the following we shall refer to this method as \emph{LoGo}.

In this paper, we demonstrate that the structure provided by   Information Filtering Networks is also extremely effective to generate high-likelihood sparse probabilistic models.
In the linear case, the {LoGo} sparse inverse covariance has only $\mathcal{O}(p)$ parameters but, despite its sparsity, the associated multivariate normal distribution can still retrieve high likelihood values yielding, \CR{in practical applications such as financial data}, comparable or better results than state-of-the-art {Glasso} penalized inversions.

\CR{The methodology introduced in this paper contributes to the literature on graphical modelling, sparse inverse covariances and machine learning. But it is of even greater relevance to physical sciences, particularly for what concerns complex systems research and network theory.
Indeed, physical sciences are increasingly engaged in the study of complex systems where the challenge is to elaborate models able to make predictions based on observational data.  
With LoGo, for the first time we combine a successful tool to describe complex systems structure (namely the  Information Filtering Networks) with a parsimonious probabilistic model that can be used for quantitative predictions. 
Such a combination is of relevance to any context where parsimonious statistical modelling are applicable.
}

The rest of the paper is organized as follows: In Section \ref{sec:methods} we describe our methodology providing algorithms for sparse covariance inversion for two graph topologies: MST and the TMFG. We then show in Section \ref{sec:results} that our method yields comparable or better results in maximum likelihood compared to lasso-type and ridge regression estimates of the inverse covariances \CR{from financial time series}. 
Subsequently we discuss how our approach can be used for time series prediction, financial stress testing and risk allocation. 
With Section \ref{sec:conclusions} we end with possible extensions for future work and conclusive remarks.
	
\begin{figure}[t]
\centering
\includegraphics[width=0.4\textwidth]{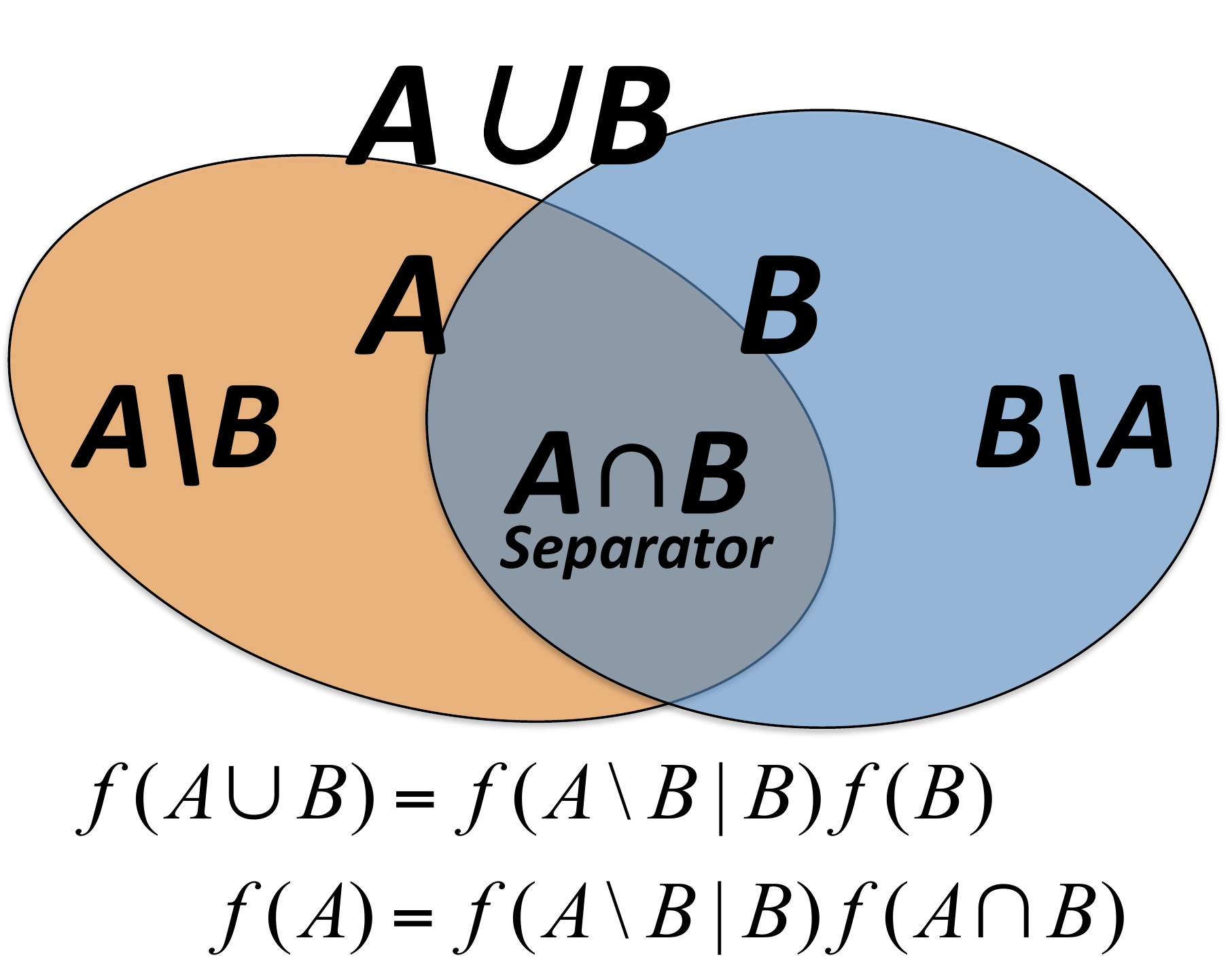}
\caption{
{\bf Decomposition of the joint probability distribution function.} Bayes formulas for two  sets of variables $A$ and $B$ with a separating set $A\cap B$.
Variables within sets $A$ or $B$ are assumed conditionally dependent whereas variables belonging to the two separated sets $A\setminus B$ and $B \setminus A$ are assumed  independent conditionally to $A\cap B$.
By combining the two formulas one obtains: $f(A\cup B) = f(A)f(B)/f(A\cap B)$.}
\label{Fig:ensambles}
\end{figure}

\section{Parsimonious modelling with  Information Filtering Networks}
\label{sec:methods}
\subsection{Factorization of the multivariate probability distribution function}
Let us start by demonstrating how a decomposable Information Filtering Network can be associated with a convenient factorization of the multivariate probability distribution.
Let us  consider, in general, two sets $A$ and $B$ of variables with non empty intersection  $A\cap B \not= \emptyset$.
Let us also assume that the variables are mutually dependent within their own ensemble $A$ or $B$  but when one variable belongs to set $A\setminus B$ and the other variable belongs to set $B \setminus A$, then they are independent conditioned to $A\cap B $.
We can now use  the Bayes formula: $f(A\cup B) = f(A\setminus B |B)f(B)$ where  $f(A \cup B)$ is the joint probability distribution function of all variable in $A$ and $B$, $ f(A\setminus B |B)$ is the conditional probability distribution function for the variables in $A$ minus the subset in common with $B$ conditioned to all variables in $B$ and  $f(B)$ is the marginal probability distribution function of all variables in $B$ (see Fig.\ref{Fig:ensambles}).
From the Bayes formula we also have the following identity: $f(A) = f(A\setminus B |B)f(A\cap B)$ that combined with the previous gives the following factorization for the joint probability distribution function of all variable in $A$ and $B$ \cite{lauritzen1996}:
\begin{equation}
f(A\cup B) = \frac{f(A)f(B)}{f(A\cap B)}
\label{eq:Bayes}
\end{equation}
Let us now apply this formula to a set of variables associated with a decomposable Information Filtering Network $\mathcal{G}$ 
made of $M_c$ cliques, ${\mathcal{C}_m}$, with $m=1,..,M_c$ and $M_s$ {complete} separators ${\mathcal{S}_n}$, with $n=1,...,M_s$.
In such a network, the vertices  {represent}  the $p$ variables $X_1,...,X_p$ and the edges    {represent} couples of \emph{conditionally} dependent variables (condition being with respect to all other variables).
Conversely, variables which are not directly connected with a network edge are conditionally independent.
Given such a network, in the same way as for Eq.\ref{eq:Bayes}, one can write the joint probability density function  for the set of $p$ variables $\bo{X}=(X_1,X_2,...,X_p)^\mathsf{T}$ in terms of the following factorization into cliques and separators  \cite{lauritzen1996}:
\begin{equation}
f(\mathbf{\bo{X}} ) = \frac{\prod_{m=1}^{M_c}
  f_{\mathcal{C}_m}(\mathbf{\bo{X}}_{\mathcal{C}_m})}{\prod_{n=1}^{M_s}
  f_{\mathcal{S}_n}(\mathbf{\bo{X}}_{\mathcal{S}_n})^{k(\mathcal{S}_n)-1}}
\label{eq:Factorizing}
\end{equation}
where $f_{\mathcal{C}_m}(\mathbf{X}_{\mathcal{C}_m})$
and $f_{\mathcal{S}_n}(\mathbf{\bo{X}}_{\mathcal{S}_n})$ are respectively the marginal probability density functions of the variables constituting $\mathcal C_m$ and $\mathcal S_n$ \cite{lauritzen1996}. The term $k(\mathcal{S}_n)$  counts the number of disconnected components produced by removing the separator $\mathcal S_n$ and it is therefore the degree of the separator in the clique tree.
Given the graph $\mathcal{G}$, Eq.\ref{eq:Factorizing} is exact, it is a direct consequence of the Bayes formula and it is therefore very general \CR{and applicable to both linear, non-linear as well as parametric or non-parametric modelling}.

\subsection{Functional form of the multivariate probability distribution function}
We search for the functional form of the multivariate probability distribution function, $f(\bo{X}  )$. 
To find the functional form of the distribution $f$ and the values of its parameters $\mathbf J$, we use the maximum entropy method \cite{Jaynes1957a,Jaynes1957b} which constrains the model to have some given expectation values while maximising the overall information entropy $ - \int f(\bo{X}  ) \log f(\bo{X}  ) d^p \bo{X} $.
At the second order, the model distribution that maximizes entropy while constraining moments at given values is:
\begin{equation} \label{MaxEnt}
f(\bo{X}  ) =
 \frac{1}{Z}\exp\left( - \sum_{ij} \frac{1}{2} (X_i-\mu_i)  J_{i,j}  (X_j-\mu_j) \right)\;\;;
\end{equation}
where $\bo{\mu}  \in\mathbb{R}^{p \times 1}$ is the vector of expectation values with coefficients $\mu_i = \E{X_i}$ and $J_{i,j}$ are the matrix elements of $\mathbf{J}\in\mathbb{R}^{p \times p}$. They are the Lagrange multipliers associated with the second moments of the distribution $\E{(X_i-\mu_i)(X_j-\mu_j)}=\Sigma_{i,j}$ which are the coefficients of the covariance matrix $\bo{\Sigma} \in\mathbb{R}^{p \times p}$ of the set of $p$ variables $\bo{X}$.
It is clear that Eq.\ref{MaxEnt} is a multivariate normal distribution with $Z = \sqrt{(2 \pi)^p \det\left( \bo{\Sigma} \right)}$.
If we require the model $f(\bo{X}  )$ to reproduce exactly all second moments $\Sigma_{i,j}$, then the solution for the distribution parameters is $\mathbf{J} = \bo{\Sigma}^{-1}$.
Therefore, in order to construct the model, one could estimate empirically the covariance matrix $ \bo{\hat \Sigma}$ from a set of $q$ observations and then invert it in order to estimate the inverse covariance.
However, in the case when the observation length $q$ is smaller than the number of variables $p$ the empirical estimate of the covariance matrix $ \bo{\hat \Sigma}$ cannot be inverted.
Furthermore, also in the case when $q>p$, such a model has $p(p+3)/2$ parameters and this might be an overfitting solution describing noise instead of the underlying relationships between the variables resulting in poor predictive power \cite{hotelling1953,schaeferstrimmer2005}. 
Indeed, we shall see in the following that, when  uncertainty is large ($q$ small), models with a smaller number of parameters can have stronger predictive power and can better describe the statistical variability of the data \cite{dempster1972}.
Here, we  consider a  parsimonious modelling that fixes only a selected number of second moments and leaves the others unconstrained.
This corresponds to model the multivariate distribution by using a {\it sparse inverse covariance} where the unconstrained moments are associated with zero coefficients in the inverse.
Let us note that this in turns implies zero partial correlation between the corresponding couples of variables.


\subsection{Sparse inverse covariance from decomposable  Information Filtering Networks}

From  Eq.\ref{eq:Factorizing} it follows that, in the case of the multivariate normal distribution, the network $\mathcal{G}$ coincides with the structure of non-zero coefficients, ${ J}_{i,j}$ in Eq.\ref{MaxEnt} and  their values can be computed from the local inversions of the covariance  matrices respectively associated with the cliques and separators  \cite{lauritzen1996}:
\begin{equation}\label{eq:locaInversion}
{ J}_{i,j} = \sum_{\mathcal{C} \; s.t.\; \{i,j\}\in\mathcal{C}} 
\left(\bo{\Sigma}_{\mathcal{C}}^{-1}\right)_{i,j}  - \sum_{\mathcal{S}  \; s.t.\; \{i,j\}\in\mathcal{S}}\left(k(\mathcal{S})-1\right) \left(\bo{\Sigma}_{\mathcal{S}}^{-1}\right)_{i,j}
\end{equation}
and ${ J}_{i,j} =0$ if $\{i,j\}$ are not both part of a common clique.\\

\begin{figure}[h!]
Planar graph made of 4-cliques separated by 3-cliques: \\
\includegraphics[width=0.2\linewidth]{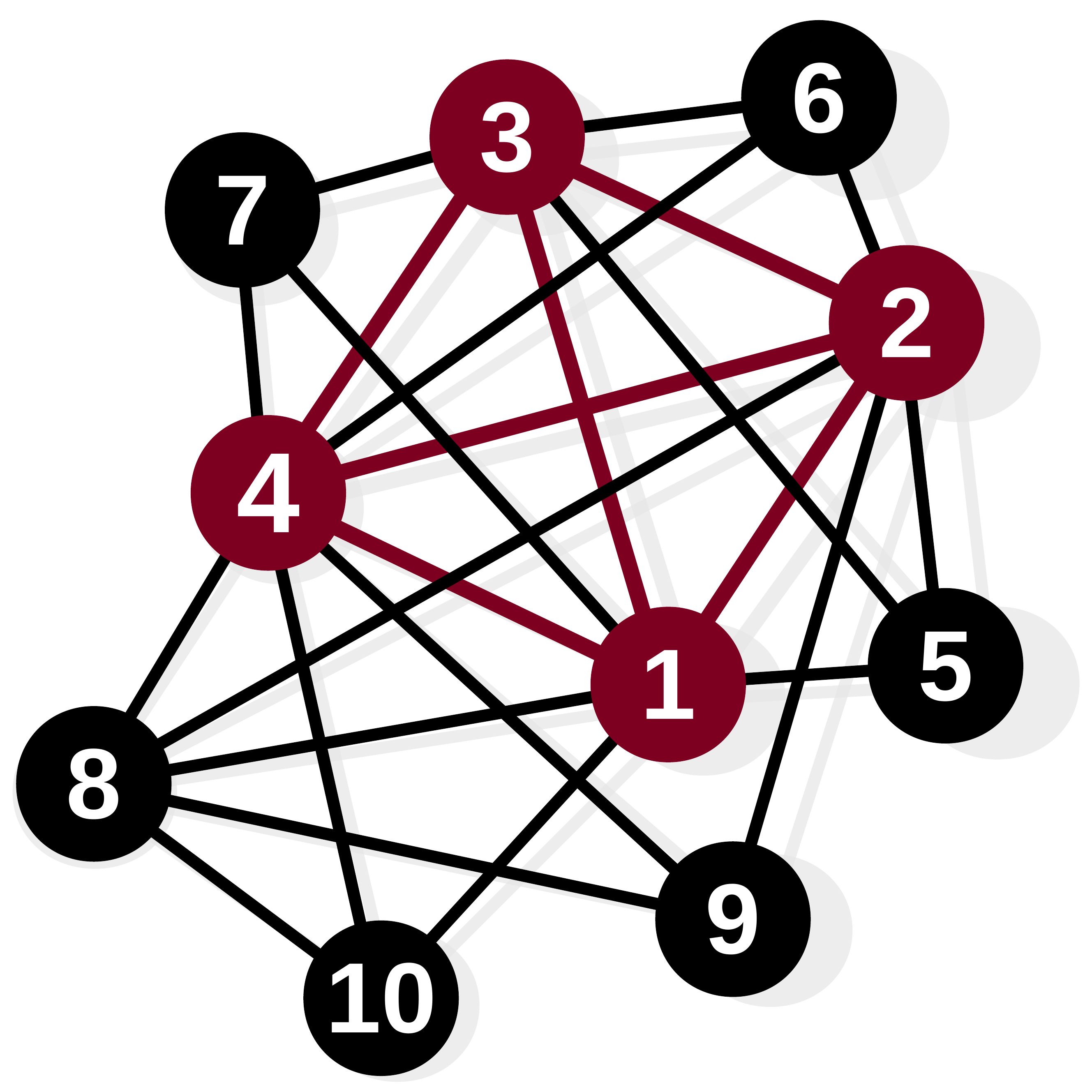}
\includegraphics[width=0.5\linewidth]{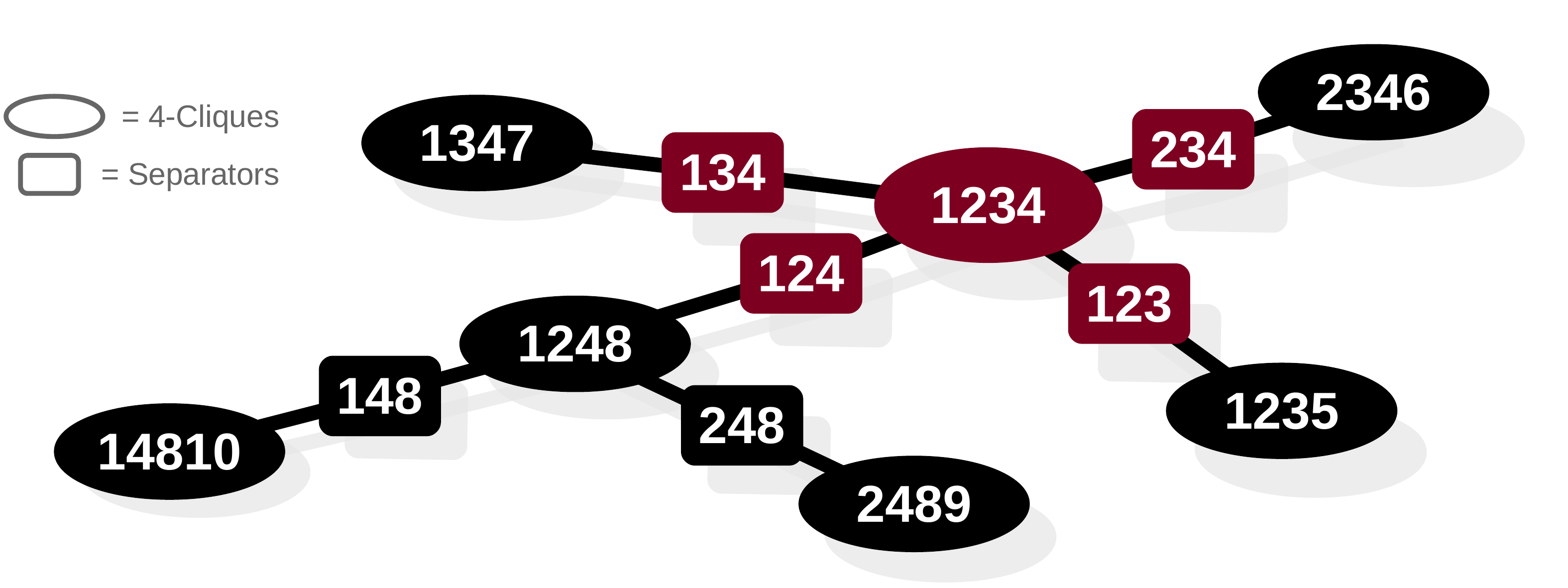}\\
Example for the computation of element $(4,8)$ of the inverse covariance:
\begin{eqnarray}
{ J}_{4,8} = 
\bo{\Sigma}^{-1}_{{\mathcal{C}}=\{1,4,8,10\}}  
 \!\! + \bo{\Sigma}^{-1}_{{\mathcal{C}}=\{1,2,4,8\}}  
  \!\! + \bo{\Sigma}^{-1}_{{\mathcal{C}}=\{2,4,8,9\}}  - \bo{\Sigma}^{-1}_{{\mathcal{S}}=\{1,4,8\}}  
- \bo{\Sigma}^{-1}_{{\mathcal{S}}=\{2,4,8\}}  \nonumber
\end{eqnarray}
\caption{
{\bf Local-Global inversion of the covariance matrix.} Example for a system of $p=10$ variables associated with a decomposable TMFG graph with $M_c=7$ cliques and $M_s=6$ separators. \label{Gdecom}}
\end{figure}

This is a very simple formula that reduces the global problem of a $p \times p$ matrix inversion into a sum of local inversions of matrices of the sizes of the cliques and separators (no more than 3 and 4 in the case of TMFG  graphs \cite{distel2010,TMFG}). 
This means that, for TMFG  graphs, only $4$ observations would be enough to produce a non-singular global estimate of the inverse covariance.
An example illustrating this inversion procedure is provided in Fig.\ref{Gdecom}.

\subsection{Construction of the maximum likelihood network}
We are now facing two related problems: 
1) How to choose the moments to retain? 
2) How to verify that the parsimonious model is describing well the statistical properties of the system of variables?
The solutions of these two problems are related because we aim to develop a methodology that chooses the non-zero elements of the inverse covariance in such a way as to best model the statistical properties of the real system under observation.
In order to construct a model that is closest to the real phenomenon we search for the set of parameters,  $\mathbf J$, associated with the largest likelihood, i.e. with the largest probability of  observing the actual observations: $\{x_{1,1},...,x_{1,q}\}$, $\{x_{2,1},...,x_{2,q}\}$....$\{x_{p,1},...,x_{p,q}\}$.
The logarithm of the likelihood 
from a model distribution function, $f(\mathbf{\bo{X}} )$ (Eq.\ref{MaxEnt}), with parameters  $\mathbf J$, is associated to the empirical estimate of the covariance matrix, $\mathbf{\hat \Sigma}$, by \cite{hoel1954introduction}: 
\begin{equation}
\ln \mathcal{L}(\mathbf J) =  
	\frac{q}{2}\left( \ln\det\mathbf J
	- \operatorname{Tr} \left( \mathbf{\hat \Sigma}\mathbf J \right)
	- p \ln(2\pi) \right).
\label{eq:GaussianLogLikelihood}
\end{equation}

The network  $\mathcal{G}$ that we aim to discover must be associated with largest log-likelihood and it can be  constructed in a greedy way by adding in subsequent steps elements with maximal log-likelihood.
In this paper we propose two constructions:
1) the maximum spanning tree (MST) \cite{kruskal1956}, which builds a spanning tree which maximises the sum of edge weights; 
2) a variant of the TMFG \cite{TMFG}, which builds a planar graph that aims to maximize the sum of edge weights.
In both cases edge weights are associated with the log-likelihood.
One can show that for all decomposable graphs, following Eq. \ref{eq:locaInversion}, the middle term in Eq. \ref{eq:GaussianLogLikelihood} is:  $\operatorname{Tr}(\mathbf{\hat \Sigma J}) = p$. 
Hence,  to maximize log-likelihood, only $\log \det\mathbf{J}$ must be maximized; from Eq.\ref{eq:Factorizing}, this is \cite{lauritzen1996}: 

\begin{equation}
 \log \det\mathbf J = 
 \sum_{n=1}^{M_s}\left(k(\mathcal{S}_n)-1\right) \log   \det\mathbf {\hat \Sigma}_{\mathcal{S}_n}
 -
 \sum_{m=1}^{M_c} \log \det\mathbf {\hat \Sigma}_{\mathcal{C}_m}
\;\; .
\label{eq:DetLikelihoodLogo}
\end{equation}

For the LoGo-MST, the construction is simplified because in a tree the cliques are the edges $e=(i,j)$, the separators are the non-leaf vertices $v_i$ and $k(\mathcal{S}_i)=k_i$ are the vertex degrees.
In this case Eq.\ref{eq:DetLikelihoodLogo} becomes
$\log \det \mathbf J = \sum_{i=1}^p \log \hat \sigma_i^{2(k_i-1)} - \sum_{e \in MST}  \log \det \hat{\bo{\Sigma}}_{e}$, with  $\hat \sigma^2_i$ the sample variance of variable `$i$'. 
Given that $\det \mathbf{\hat \Sigma}_{e} = \hat \sigma^2_i \hat \sigma^2_j(1-\hat R_{i,j}^2)$, with $\hat R_{i,j}$ the Pearson's correlation matrix element $i,j$, then  $\log\det \mathbf J = - \sum_{i=1}^p \log\hat\sigma_i^2 - \sum_{e \in MST} \log (1- \hat R_{i,j}^2)$.
Therefore,  the MST can be built through the standard Prim's \cite{prim1957} or  Kruskal's \cite{kruskal1956} algorithms from a matrix of weights $\mathbf W$ with coefficients $W_{i,j}=\hat R_{i,j}^2$.
The \emph{LoGo-MST} inverse covariance estimation, $\mathbf{J}$, is then computed by the local inversion, Eq.\ref{eq:locaInversion}, on the MST structure.
Note that  the MST  structure depends only on the correlations not the covariance.

\begin{figure}[t]
\centering
\includegraphics[width=0.7\textwidth]{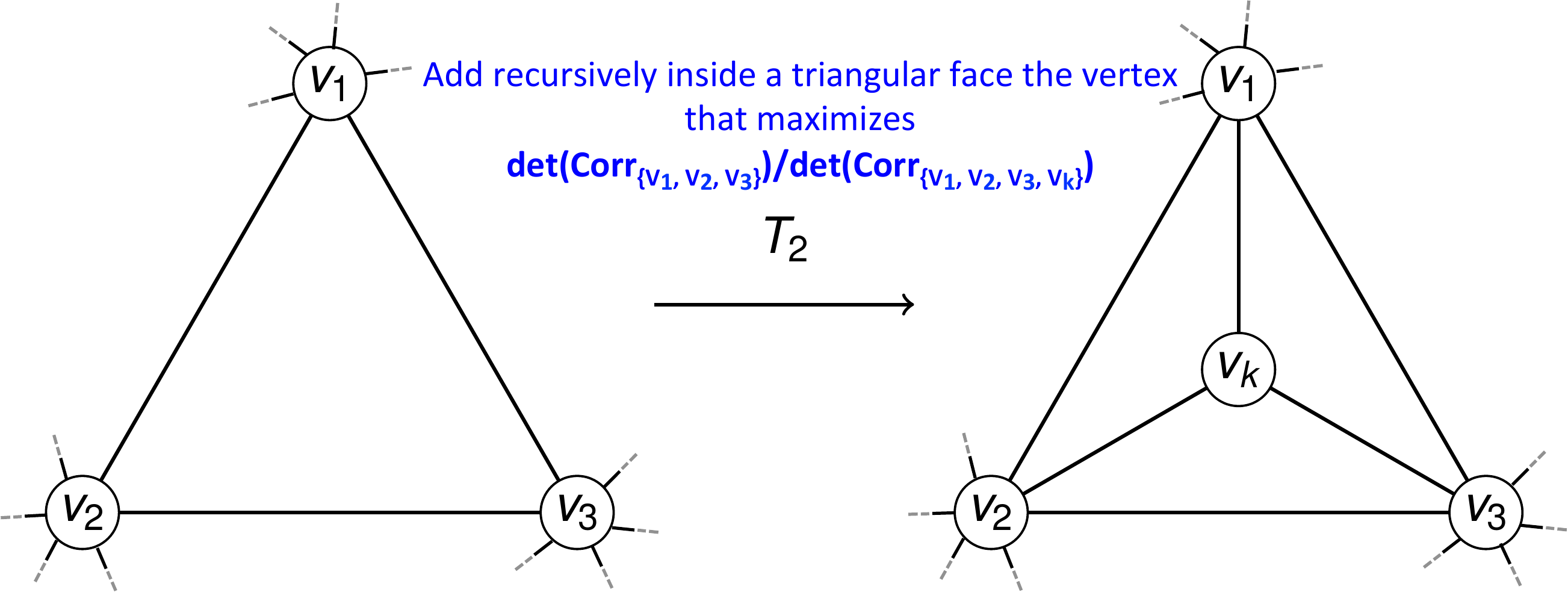}
\caption{
{\bf TMFG construction.} The TMFG graph is generated by adding vertices (e.g. $v_k$) inside triangular faces (e.g. $\{v_1,v_2,v_3\}$) maximising the ratio of the determinants between separator and clique $\det(\mathbf{\hat R}_{\{v_1,v_2,v_3\}})/\det(\mathbf{\hat R}_{\{v_1,v_2,v_3,v_k\}})$.
This move generates a new 4-clique $\mathcal C = \{v_1,v_2,v_3,v_k\}$ and transforms the triangular face into a separator $\mathcal S = \{v_1,v_2,v_3\}$.}
\label{fig:TMFG}
\end{figure}

The LoGo-TMFG construction requires a specifically designed procedure. Also in this case, only correlations matter; indeed, the  structure of the inverse covariance network reflects the partial correlations i.e. the correlation between two variables given all others. 
LoGo-TMFG starts with a tetrahedron, $\mathcal{C}_1=\{v_1,v_2,v_3,v_4\}$, with smallest correlation determinant $ \det \mathbf{\hat R}_{\mathcal{C}_1}$  and then iteratively introduces inside existing triangular faces the vertex that maximizes  $\log   \det\mathbf {\hat R}_{\mathcal{S}}-\log \det\mathbf {\hat R}_{\mathcal{C}}$ where ${\mathcal{C}}$ and ${\mathcal{S}}$ are the new clique and separator created by the vertex insertion. 
The \emph{LoGo-TMFG} procedure is schematically reported in  Algorithm \ref{alg:4cliqueTree} and in Fig.~\ref{fig:TMFG}.
The TMFG is a computationally efficient algorithm \cite{TMFG} that produces a decomposable (chordal) graph, with $3(p-2)$ edges, which is a clique-tree constituted by four-cliques connected with three-cliques separators. 
Note that for  TMFG  $k(\mathcal{S}_n)=2$ always.

\SetNlSkip{1em}
\SetInd{0.15em}{0.45em}
\begin{algorithm}[h] 
	\SetKwData{VertexList}{$\mathcal{V}$}
	\SetKwData{CurrentFaces}{CurrentFaces}
	\SetKwData{Gains}{Gains}
	\SetKwData{MaxGain}{MaxGain}
	\SetKwData{BestVertex}{BestVertex}
	\SetKwData{W}{W}
	\SetKwData{TMFG}{TMFG}
	\SetKwData{tetra}{$th_1$} 	
	\SetKwData{ta}{$t_1$} 	
	\SetKwData{tb}{$t_2$} 	
	\SetKwData{tc}{$t_3$} 	
	\SetKwData{td}{$t_4$}
	\SetKwData{Triangles}{$\mathcal{T}$}
	\SetKwData{Separators}{$S$}
	\SetKwInOut{Input}{input}
	\SetKwInOut{Output}{output}
	\Input{A covariance matrix $\mathbf {\hat \Sigma} \in \mathbb{R}^{p\times p}$ and associated correlation matrix $\mathbf {\hat R} \in \mathbb{R}^{p\times p}$ from a set of observations $\{x_{1,1},...,x_{1,q}$\},  \{$x_{2,1},...,x_{2,q}$\}.... \{$x_{p,1},...,x_{p,q}$\} }
	\Output{ $\mathbf { J}$ a sparse estimation of $\mathbf { \Sigma}^{-1}$ }
	${\mathbf J}\leftarrow \mathbf 0$ Initialize ${\mathbf J}$ with zero elements\;
	$\mathcal{C}_1\leftarrow$ Tetrahedron, $\{v_1,v_2,v_3,v_4\}$, with smallest $ \det \mathbf{\hat R}_{\mathcal{C}_1}$ \;
	$\Triangles  \leftarrow$ Assign to $\Triangles$ the four triangular faces in $\mathcal{C}_1$: $\{v_1,v_2,v_3\}$,  $\{v_1,v_2,v_4\}$,  $\{v_1,v_4,v_3\}$,  $\{v_4,v_2,v_3\}$  \;
	$ \VertexList \leftarrow$ Assign to \VertexList the remaining $p-4$ vertices not in $\mathcal{C}_1$  \;
	\While{\VertexList is not empty}{
		find the combination of  $\{v_a,v_b,v_c\} \in \Triangles$ and $v_d \in \VertexList$ with largest $\det(\mathbf{\hat R}_{\{v_a,v_b,v_c\}})/\det(\mathbf{\hat R}_{\{v_a,v_b,v_c,v_d\}})$\;
		\tcp{$\{v_a,v_b,v_c,v_d\}$ is a new 4-clique $\mathcal{C}$, $\{v_a,v_b,v_c\}$ becomes a separator $\mathcal{S}$, three new triangular faces, $\{v_a,v_b,v_d\}$, $\{v_a,v_c,v_d\}$ and $\{v_b,v_c,v_d\}$ are created}
		Remove $v_d$ from \VertexList\;
		Remove $\{v_a,v_b,v_c\}$ from \Triangles \;
		Add $\{v_a,v_b,v_d\}$, $\{v_a,v_c,v_d\}$, $\{v_b,v_c,v_d\}$ to \Triangles \;
		Compute ${ J}_{i,j} = { J}_{i,j}  + \left(\bo{\Sigma}_{ \{v_a,v_b,v_c,v_d\} }^{-1}\right)_{i,j}  -  \left(\bo{\Sigma}_{\{v_a,v_b,v_c\}}^{-1}\right)_{i,j}$\;
	}
	\Return $\mathbf{ J}$\;
	\caption{{\bf LoGo-TMFG algorithm.} Construction of sparse LoGo-TMFG sparse inverse covariance $\mathbf { J}$ starting from a covariance matrix  $\mathbf {\hat \Sigma}$. The non-zero elements of $\mathbf { J}$ are the edges of $\mathcal{G}$ which is a chordal graph, a clique tree, made of tetrahedra separated by triangles.} \label{alg:4cliqueTree}
\end{algorithm}%

Let us note that by expanding to the second order in the correlation coefficients, the logarithms of the determinants are well approximated by  a constant minus the sum of the square correlation coefficients associated with the edges in the cliques or separators. 
This can simplify the algorithm and the TMFG could be simply computed from a set of weights given by the squared correlation coefficients matrix, as described in \cite{TMFG}.

Further, let us note that, for simplicity, in this paper we only consider likelihood maximization. A natural, straightforward extension of the present work is to consider Akaike Information Criterion \cite{akaike1974new,akaike1998information} instead. However, we have verified that, for the cases studied in this paper the two approaches give very similar results. 

\begin{figure}[t]
\centering
\includegraphics[width=0.65\textwidth]{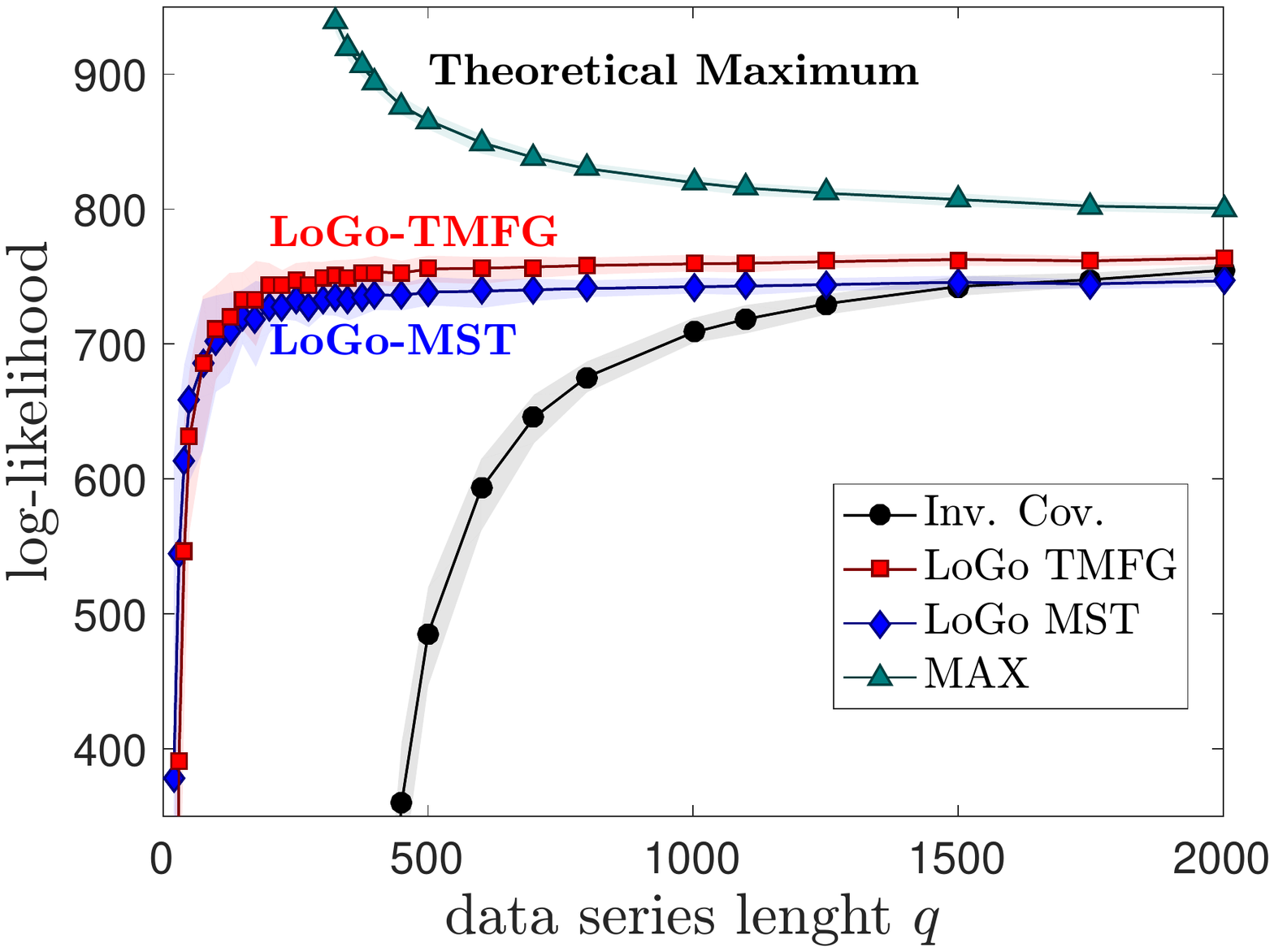}
\caption{
{\bf Demonstration that LoGo sparse inverse covariance represents the dependency structure better than the complete inverse covariance.}
The figure reports comparisons between log-likelihood for models constructed by using sparse inverse LoGo-TMFG, LoGo-MST and the complete inverse  of the empirical covariance matrix (Inv. Cov.). 
These measures are on $p=300$ off-sample test data-series of different lengths $q$ varying from 20 to 2000.
The inverse matrices are computed on training datasets of the same length.
Data are log-returns sampled from 342 stocks prices of equities traded on the US market during the period 1997-2012.
The statistics is made stationary by random shuffling the time order. 
Symbols correspond to averages over 100 samples generated by picking at random 300 series over the 342 and assigning training and testing sets by choosing at random two non-overlapping time-windows of length $q$, the shaded bands are the 95\% quantiles.
The line on the top, labelled with `MAX', is the theoretical maximum which is the  log-likelihood obtained from the inverse covariance of the testing set. 
}
\label{fig:LL_comparisonWithInverse}
\end{figure}

\begin{figure}[t]
\centering
\includegraphics[width=0.65\textwidth]{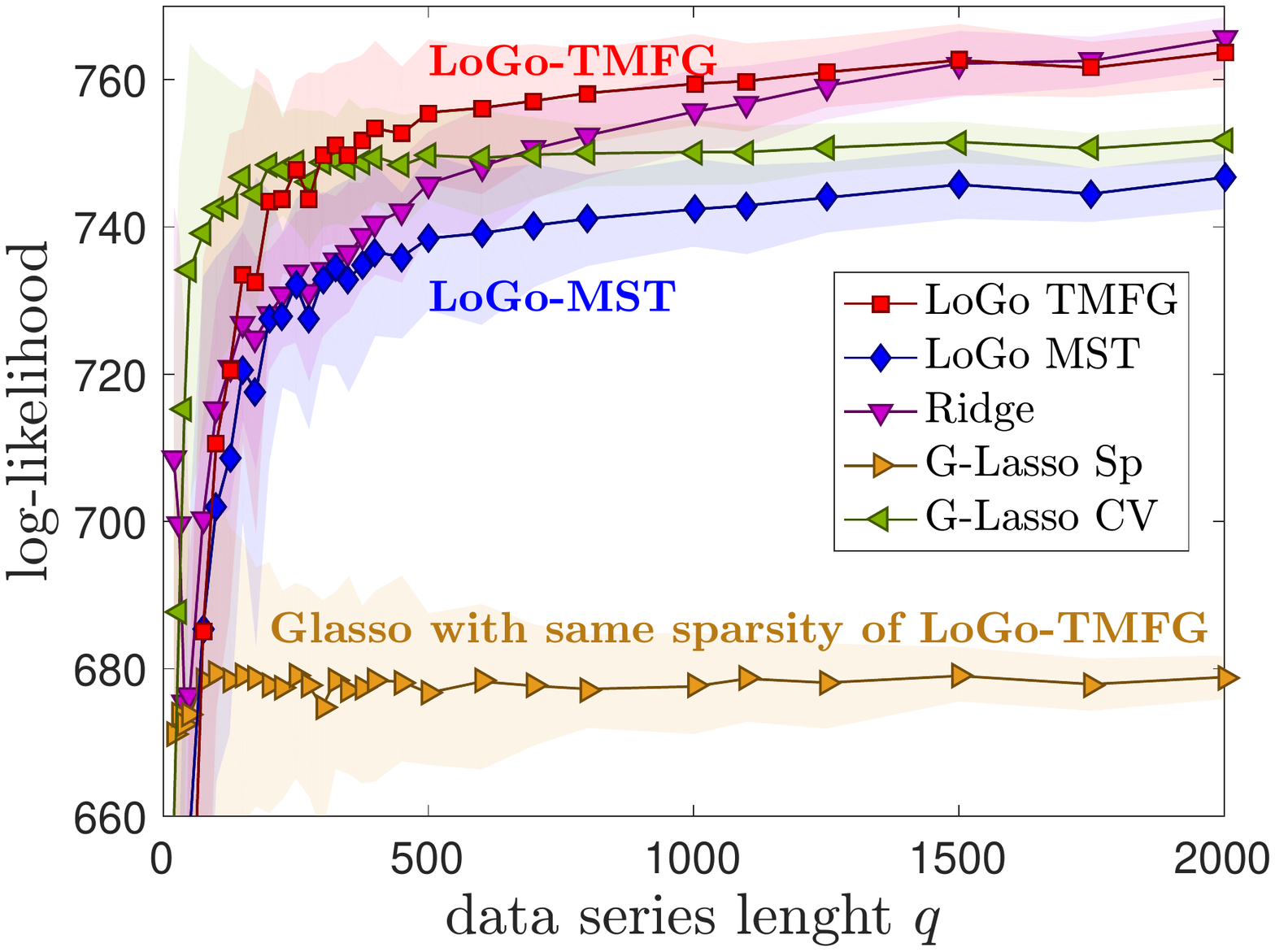}
\caption{
{\bf Demonstration that LoGo sparse modelling has better performances than Glasso models with same sparsity.}
This figure reports the same log-likelihood as in Fig.\ref{fig:LL_comparisonWithInverse}  compared with  log-likelihood  from state-of-the-art Glasso $\ell_1$ penalized sparse inverse covariance models (cross validated, G-Lasso-CV, and of the same sparsity of TMFG, G-Lasso-Sp) and Ridge $\ell_2$ penalized inverse model (Ridge). 
}
\label{fig:LL_comparisonWithGlasso}
\end{figure}

\begin{figure}[t]
\centering
\includegraphics[width=0.65\textwidth]{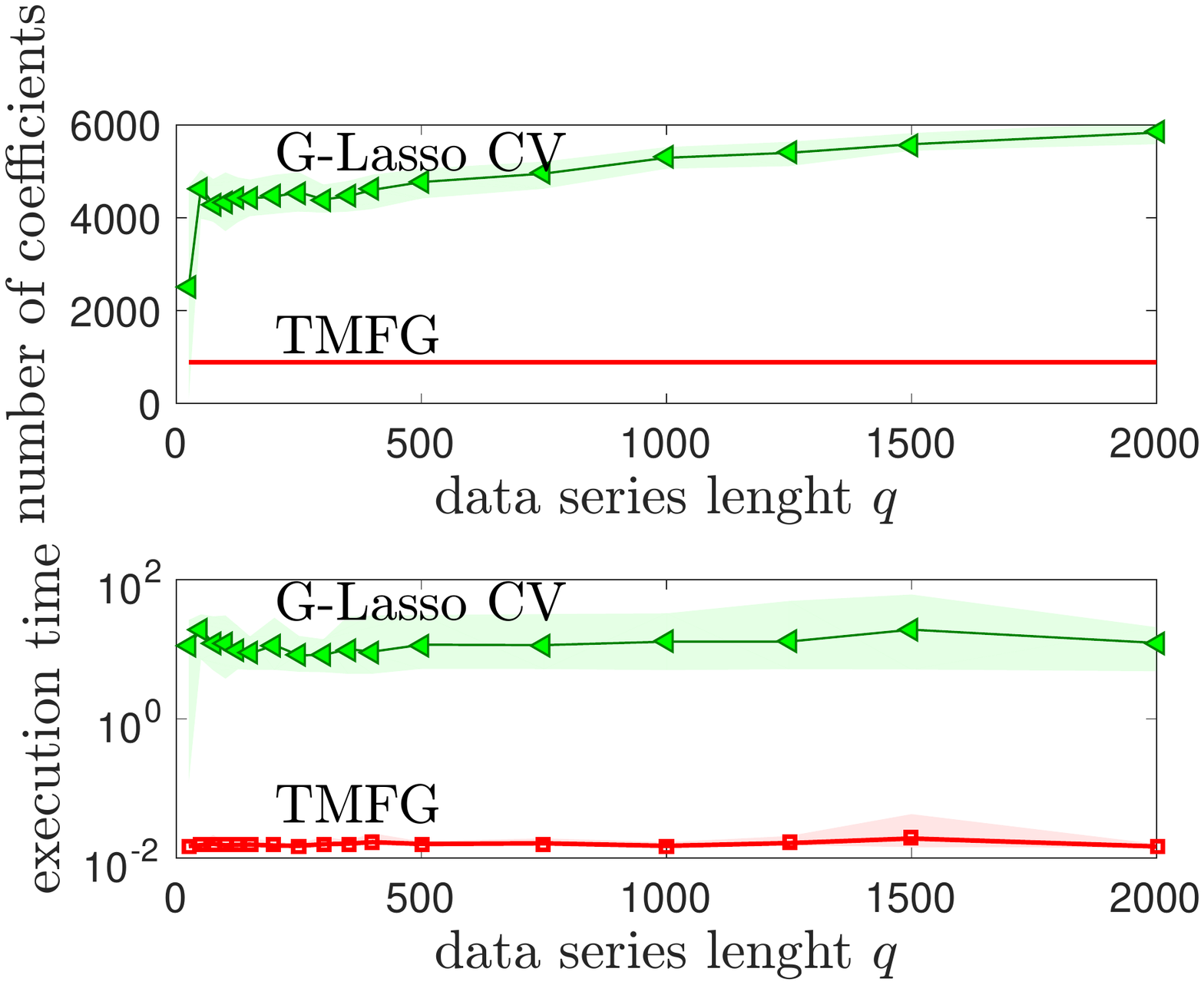}
\caption{
\CR{{\bf Demonstration that LoGo  models are sparser than Glasso models and are computationally more efficient.}
The plot on the top reports the number of non-zero off-diagonal coefficients in the precision matrix $\mathbf J$ for G-Lasso-CV and LoGo-TMFG.
The plot on the bottom reports the computation times (seconds) for G-Lasso-CV and LoGo-TMFG.
These data refer to the same simulations as for the results in Figs.\ref{fig:LL_comparisonWithInverse} and \ref{fig:LL_comparisonWithGlasso}. Note that TMFG-LoGo has a constant number of coefficients equal to $3(p-2) = 894$.}}
\label{fig:LL_comparisonWithGlasso_spersityAndTime}
\end{figure}

\begin{figure}[t]
\centering
\includegraphics[width=0.65\textwidth]{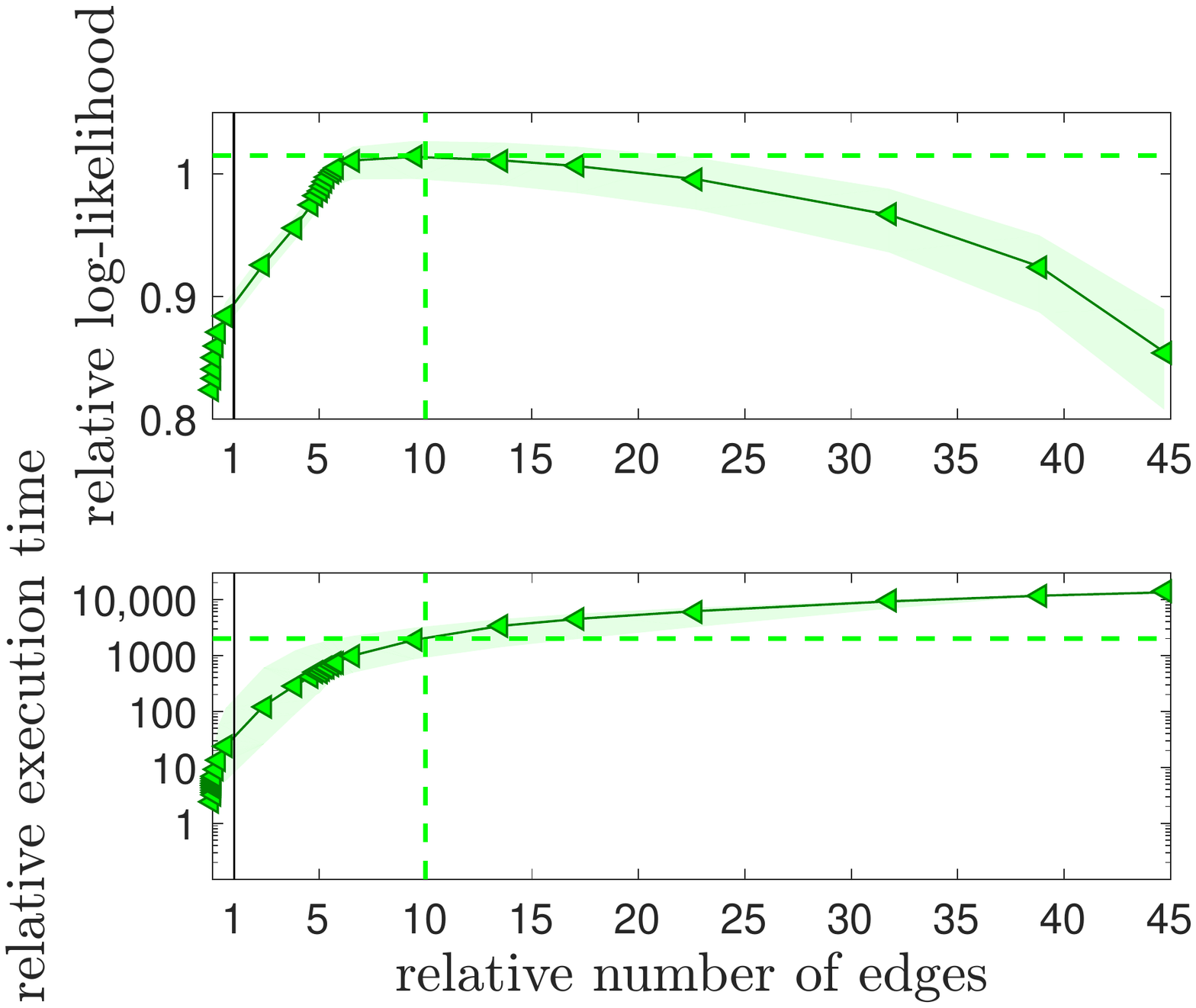}
\caption{
{\bf Demonstration that LoGo sparse models with comparable performances with respect to best performing Glasso methods produces sparser models in a fraction of the computation time.}
The top plot reports  the fraction of log-likelihood for Glasso and for LoGo-TMFG vs. the fraction of non zero off-diagonal coefficients of $\mathbf{J}$ for  Glasso and for LoGo-TMFG.
The bottom plot reports the fraction of computational time  vs. the fraction of non zero off-diagonal coefficients of $\mathbf{J}$ for  Glasso and for LoGo-TMFG.
The measures are on $p=300$ off-sample test data-series of  length $q=500$.
Inverse matrices are computed on training datasets of the same length.
Data are log-returns sampled from 342 stocks prices of equities traded on the US market during the period 1997-2012.
The statistics is made stationary by random shuffling the time order. 
Symbols correspond to averages over 100 samples generated by picking at random 300 series over the 342 and assigning training and testing sets by choosing at random two non-overlapping time-windows of length $q$, the shaded bands are the 95\% quantiles.
}
\label{fig:LL_comparisonWithGlasso_sparsityAndTime2}\label{fig.S1}
\end{figure}

\section{Results}
\label{sec:results}
\subsection{Inverse covariance estimation}
We investigated stock prices time series from a US equity market computing the daily log-returns ($x_i(t) = \log Price_i(t) - \log Price_i(t-1)$ with $i=1,..,342$ and $t=1,...,T$ with $T=4025$ days during a period of 15 years from 1997 to 2012 \cite{musmeci2015relation}).
We build 100 different datasets by creating stationary time series of different lengths selecting returns at random points in time and randomly picking $p=300$ series out of the 342 in total.
Each dataset was divided into two temporal non-overlapping windows with $q$ elements constituting  the `training set' and other $q$ elements the `testing set'.

In order to quantify the goodness of the methodology we  computed the log likelihood of the testing dataset using the inverse covariance estimates  from the training set.  
Larger log likelihood indicate models that better describe the testing data.
Figure \ref{fig:LL_comparisonWithInverse} reports the results for time series of different lengths from $q=25$ to $q=2000$.
\CR{Smaller values of $q$ mean shorter number of observations in the training dataset used to construct the model and therefore correspond to larger uncertainties.}
Note that, the green upward triangles in Fig.~\ref{fig:LL_comparisonWithInverse}, denoted with \emph{MAX}, are the theoretical maximum from the inverse sample covariance matrix calculated on the testing set which is reported as a reference indicating the upper value for the attainable likelihood.
Let us first observe from this figure that, \CR{for these stationary financial time series study,} \emph{LoGo-TMFG}  outperforms the likelihood from the inverse covariance solution $\mathbf{J=\hat \Sigma}^{-1}$ (denoted with `Inv. Cov.' in Fig.~\ref{fig:LL_comparisonWithInverse}). 
For $q<p=300$ the  inverse covariance is not computable and therefore comparison cannot be made; when $q>p=300$, the inverse covariance is computable but it performs very poorly for small sample sizes  $q\sim p$ becoming comparable to \emph{LoGo-TMFG} only after $q\sim 1500$ 
with both approaching the theoretical maxima at $q\rightarrow \infty$.
Note that also \emph{LoGo-MST} outperforms the inverse covariance solution in most of the range of $q$.
We then compared the log-likelihood from \emph{LoGo-MST} and \emph{LoGo-TMFG} sparse inverse covariance with state-of-the-art Glasso $\ell_1$-norm penalized sparse inverse covariance models (using the implementation by \cite{hsieh2014quic}) and Ridge $\ell_2$-norm penalized inverse model. 
Glasso method depends on the regularization parameters which were estimated by using two standard methods: 
i) \emph{G-Lasso-CV}  uses a two-fold cross validation method  \cite{pedregosaetal2011}; 
ii)  \emph{G-Lasso-Sp}  fixes the regularization parameter to the value that creates in the training set a sparse inverse with sparsity equal to \emph{LoGo-TMFG} network ($3(p-2)$ parameters).
\emph{Ridge} inverse penalization parameter was also computed by cross validation method  \cite{pedregosaetal2011}. 
Fig.\ref{fig:LL_comparisonWithGlasso} reports a comparison between these methods for various values of $q$.
We can observe that \emph{LoGo-TMFG} outperforms the Glasso methods achieving larger  likelihood from  $q>100$. 
Results are detailed in Table~\ref{tab:RealDataShuff} where we compare also with the null model (`NULL') which is a completely disconnected network corresponding to a diagonal $\mathbf J$.

LoGo models can achieve better performances than Glasso models with fewer coefficients and are computationally more efficient. 
This is shown in  Fig.\ref{fig:LL_comparisonWithGlasso_spersityAndTime}  where we report the comparison between the number of non-zero off-diagonal coefficients in the precision matrix $\mathbf J$  in Glasso-CV and LoGo models. These results shows that the number of coefficients  for G-Lasso-CV is 3 to 6 times larger than for LoGo-TMFG  while the computation time for LoGo-TMFG is about three order of magnitude smaller than the computation time for G-Lasso-CV.
Note that LoGo-TMFG has a constant number of  coefficients equal to $3p-6$ corresponding to the number of edges in the TMFG network.
A further comparison between performance, sparsity and execution time is provided in Fig.~\ref{fig:LL_comparisonWithGlasso_sparsityAndTime2} where the top plot reports the fraction of log-likelihood for Glasso (implemented by using \cite{hsieh2014quic}) and for LoGo-TMFG vs. the fraction of non zero off-diagonal coefficients of $\mathbf{J}$  for data series lengths of $q=500$.  
In the bottom plot we report instead the fraction of computational time  vs. the fraction of non zero off-diagonal coefficients of $\mathbf{J}$ for  Glasso and for LoGo-TMFG.
We can observe that at the same sparsity (value 1 in the x-axis indicated with the vertical line)  Glasso underperforms LoGo by 10\% and Glasso is about 50 times slower than LoGo on the same machine.
We verified that eventually the maximum performance of Glasso can become 1.5\% better than LoGo but with 10 times more parameters and computation time 2,000 times longer.
Let us note that, in this example, the best Glasso performance are measured a-posteriori on the testing set,  they are therefore hypothetical maxima which cannot be reached with cross validation methods that instead result in average performances of a few percent inferior to LoGo (see Fig.\ref{fig:LL_comparisonWithGlasso}).
All these results refer to the same simulations as for the results in Figs.\ref{fig:LL_comparisonWithInverse}.
\wolf{The computation time of LoGo can be decreased even further by parallelising the algorithm.}


The previous results are for time series made stationary by random selecting log-returns at different times. 
In practice, financial time series - and other real world signals - are non-stationary having statistical properties that change with time.
 Table~\ref{tab:RealData} reports the same analysis as in Fig.\ref{fig:LL_comparisonWithGlasso} and Table~\ref{tab:RealDataShuff} but with datasets taken in temporal sequence with the training set being the $q$ data points preceding the testing set.
Let us note that, considering the time-period analysed, in the case of large $q$, the training set has most data points in the period preceding  the 2007-2008 financial crisis whereas the testing set has data in the period following the crisis.
Nonetheless, we see that the results are comparable with the one obtained for the stationary case. 
Surprisingly, we observe that for relatively small time-series lengths the values of the log-likelihood achieved by the various models is  larger than in the stationary case.
This counter-intuitive fact can be explained by the larger temporal persistence of  real data with respect to the randomized series.
\CR{Further results and a plot of the log likelihood for this non stationary case are given in Appendix \ref{Sii}  (Fig.\ref{fig.S2}).}

We also investigated  artificial datasets of $p=300$ multivariate variables generated from factor models respectively with 3 and 30 common factors.
Results for the average log-likelihood and the standard deviations computed over 100 samples at different values of $q$ are reported in Table~\ref{tab:factorModels}.
\CR{In this case we note that, while for a number of factors equal to 3  LoGo is again performing consistently better than the inverse covariance, better than Glasso models of equal sparsity and comparably well with cross validated Glasso; conversely, when the number of factors is set to 30 LoGo is underperforming with respect to cross validated Glasso and even the inverse covariance (for $q>400$).
However, we note that LoGo is doing better than the Glasso model with equal sparsity.
This seems to indicate that factor models with more than a few factors require denser models with larger numbers of non zero elements in the inverse covariance than the one provided by MST and TMFG networks used in the present LoGo construction. }
Note that by increasing the number of factors performances of all models become worse.
\CR{Let us finally note that ridge inverse covariance performs  well in all the cases studied. It is indeed well known that this is a powerful estimator for the inverse covariance, however the purpose of the present investigation concerns sparse modelling and ridge inverse covariance is dense with all coefficients different from zero. Furthermore, let us remark that LoGo can outperform ridge in several cases, as we can notice from Fig.\ref{fig:LL_comparisonWithGlasso} and Tabs.\ref{tab:RealDataShuff}, \ref{tab:RealData} and \ref{tab:factorModels}.}
\CR{Further results and a plot of the log likelihood for  factor models are given in Appendix \ref{Siii}  (Fig.\ref{fig.S3}).}

\begin{table}
\caption{\label{tab:RealDataShuff}
{\bf Demonstration that LoGo sparse modelling has better or comparable performances that state-of-the-art models.}
Comparison between log-likelihood for LoGo, Glasso, Ridge, Complete inverse and Null models. 
Measures are on $p=300$ off-sample test data-series of lengths $q=50, 100, 300, 500, 1000, 2000$.
Data are the same as in Fig.\ref{fig:LL_comparisonWithGlasso}: log-returns sampled from 342 stocks prices made stationary by random shuffling the time order in the time-series. 
The values reported are the averages of 100 samples and the standard deviations are reported between brackets. 
`MAX', is the theoretical maximum  log-likelihood obtained from the inverse covariance of the testing set. 
}
\centering
\begin{tabular}{lllllll}
\hline
$q$ &50&100&300&500&1000&2000\\
\hline
Inv. Cov. 	&  - &  - &  112 (53) &  485 (29) &  709 (8) &  755 (4) \\
LoGo TMFG 	&  631 (52) &  711 (28) &  753 (9) &  756 (8) &  759 (5) &  764 (3) \\
LoGo MST 	&  658 (43) &  702 (28) &  737 (9) &  738 (8) &  742 (5) &  747 (3) \\
Ridge     	&  676 (17) &  715 (11) &  741 (6) &  746 (6) &  756 (5) &  766 (3) \\
G-Lasso Sp 	&  674 (23) &  679 (18) &  678 (10) &  677 (8) &  678 (5) &  679 (2) \\
G-Lasso CV 	&  734 (25) &  743 (15) &  750 (5) &  750 (5) &  750 (4) &  752 (2) \\
MAX      	&  - &  - &  895 (6) &  866 (5) &  820 (4) &  801 (3) \\
NULL        	&  -276 (0.02) &  -276 (0.02) &  -276 (0.01) &  -276 (0.01) &  -276 (0.01) &  -276 (0.00) \\
\hline
\end{tabular}
\end{table}

\begin{table}
\caption{\label{tab:RealData}
{\bf LoGo performances on historic data.}
Same analysis as in Table~\ref{tab:RealDataShuff} performed for historic data where the training sets are past log-returns and the testing sets are future log-returns from two non-overlapping adjacent windows of length $q$.
}
\centering
\begin{tabular}{lllllll}
\hline
$q$ &50&100&300&500&1000&2000\\
\hline
Inv. Cov. 	&  - &  - &  - &  253 (835) &  481 (335) &  678 (15) \\
LoGo TMFG 	&  679 (212) &  757 (120) &  674 (247) &  665 (312) &  605 (211) &  694 (12) \\
LoGo MST 	&  699 (202) &  747 (120) &  656 (255) &  641 (324) &  583 (221) &  678 (11) \\
Ridge     	&  753 (113) &  746 (86) &  721 (113) &  721 (155) &  684 (115) &  710 (10) \\
G-Lasso Sp 	&  722 (110) &  704 (107) &  664 (113) &  627 (159) &  625 (97) &  659 (4) \\
G-Lasso CV 	&  769 (134) &  761 (89) &  718 (110) &  700 (178) &  666 (120) &  711 (6) \\
MAX      	&  - &  - &  - &  919 (63) &  869 (40) &  851 (3) \\
NULL        	&  -276 (0.14) &  -276 (0.07) &  -276 (0.08) &  -276 (0.07) &  -276 (0.06) &  -276 (0.00) \\
\hline
\end{tabular}
\end{table}

\begin{table}
\caption{\label{tab:factorModels}
{\bf LoGo performances on artificial data.}
Same analysis as in Table~\ref{tab:RealDataShuff} performed for artificial data generated with two factor models respectively with  $3$ (a) and $30$ (b) factors.
}
\centering
\begin{tabular}{lllll}
(a)\\
\hline
$q$ &50&400&1000&2000\\
\hline
Inv. Cov. 	& - &	-394 (14) &	-63 (17) &	-46 (18) \\	
LoGo TMFG &	-98 (13) &	-51 (17) &	-51 (18) &	-58 (20) \\	
LoGo MST &	-142 (10) &	-116 (13) &	-116 (11) &	-121 (13)\\ 	
Ridge    & 	-163 (8) &	-82 (16) &	-53 (15) &	-48 (17) 	\\
G-Lasso Sp &	-452 (9) &	-447 (4) &	-447 (3) &	-447 (2) 	\\
G-Lasso CV  &	-128 (7) &	-62 (13) &	-61 (11) &	-55 (15) 	\\
MAX     & 	- &	59 (18) &	2 (17) &	-20 (18) \\	
NULL      &  	-427 (12) &	-425 (6) &	-426 (3) &	-425 (2)\\ 	
\hline
\end{tabular}
\hskip0.5cm
\begin{tabular}{lllll}
(b)\\
\hline
$q$ &50&400&1000&2000\\
\hline
Inv. Cov. 	& - & 	-484 (12) &	-143 (12) &	-115 (11)\\ 	
LoGo TMFG 	& -440 (7) & 	-376 (2) 	& -373 (1) &	-372 (2) 	\\
LoGo MST &	-432 (5) 	& -400 (3) &	-393 (1) &	-392 (1) 	\\
Ridge  &   	-324 (11) 	& -158 (9) &	-129 (12) &	-113 (11) 	\\
G-Lasso Sp &	-430 (3) &	-425 (1) &	-424 (1) &	-424 (1)\\ 	
G-Lasso CV &	-326 (5) &	-151 (6) &	-130 (14) & 	-118 (11) \\	
MAX    &  -	& -18 (9) &	-78 (12) &	-88 (11) 	\\
NULL       & 	-425 (4) &	-426 (1) 	& -425 (1) &	-426 (1) 	\\
\hline
\end{tabular}\\

\end{table}

\subsection{ Time series prediction: \CR{LoGo for regressions and causality} }
LoGo estimates the joint probability distribution function yielding the set of parameters for the model system's dependency structure.
\CR{In this section we demonstrate how this model  can be used also for forecasting.}
\CR{ Information Filtering Networks have been proven to be powerful tools to characterize the structure of complex systems comprising several variables -such as financial markets and biological systems \cite{tumminelloetal2005,NJP10,musmeci2015relation,song2012hierarchical,musmeci2016}.
They have also been proven effective to understand how financial risk is distributed in  markets and how to construct performing portfolios  \cite{pozzi2013spread,musmeci2015risk}.
However, so far, they were not associated to probabilistic models able to make use of their capability of meaningful representation of the market structure.
LoGo provides this instrument and in particular we here show how  Information Filtering Networks can be used  to compute sparse regressions.
Indeed, generally speaking, a regression consists in estimating the expectation values of a set of variables, $\bo{X}_2$, conditioned to the values of another set of variables $\bo{X}_1$:
\begin{equation}\label{LinearReg}
\bo{\mu}_{2|1} = \mathbb E[\bo{X}_2|\bo{X}_1]\;\;.
\end{equation}
when multivariate normal statistics is used, this is the linear regression. 
If LoGo sparse inverse covariance $\mathbf{J}$ is used, then Eq.\ref{LinearReg} computes a sparse linear regression. 
If the set of variables $\bo{X}_1$ are `past' variables and  the variables $\bo{X}_2$ are `future' variables, then the regression becomes a forecasting instrument where values of variables in the  future can be inferred from past observations. 
Here we  consider to have $p_1$ variables in the past ($\bo{X}_1$) and $p_2$ variables in the future ($\bo{X}_2$ ).
These variables  can either be the same variables at different lags or different variables. 
For simplicity of notation, and without loss of generality, we consider centred variables with zero expectation values.
We can consider $\bo{X}_1$ and $\bo{X}_2$ as two distinct sets of variables that, of course, have some dependency relation. }
The conditional expectation values $\bo{\mu}_{2|1} $ can be  calculated from the conditional joint distribution function, which, from the Bayes theorem, is $f(\bo{X}_2|\bo{X}_1)=f(\bo{X}_2,\bo{X}_1)/f(\bo{X}_1)$.
From this expression one obtains:
\begin{equation}
\mathbf{J}_{2,2}\bo{\mu}_{2|1} = - \mathbf{J}_{2,1} \bo{X}_1  \;\;\;,
\label{J}
\end{equation}
where we have written the precision matrix $\mathbf{J}$ as a block matrix
\begin{equation}
 \mathbf{J}  =  \left( \begin{array}{ccc}
 \mathbf{J}_{1,1} &  \mathbf{J}_{1,2}  \\
 \mathbf{J}_{2,1} &  \mathbf{J}_{2,2}  \end{array} \right) \;\;\;
\end{equation}
whith $\mathbf{J}_{2,2}$ being the $p_2\times p_2$ part of the precision matrix in the lower right and $\mathbf{J}_{2,1}$  the $p_2\times p_1$ part of the precision matrix in the lower left. 
\CR{As pointed out previously, Eq.\ref{J} is just a} different way to write the linear regression which, in a more conventional form, reads: $\bo{X}_2 = \bo{\beta} \bo{X}_1 + \bo{\epsilon}$ with the coefficients $\bo{\beta}= -\mathbf{J}_{2,2}^{-1}\mathbf{J}_{2,1}$ and the residuals given by $\bo{\epsilon} = \bo{X}_2 -  \bo{\mu}_{2|1}$.
\CR{However, by using LoGo to estimate $\mathbf{J}$,  Eq.\ref{J} becomes a sparse predictive model associated with a meaningful inference structure. 
Now, through Eq.\ref{J}  we can quantify the effect of past values of a set of variables over the future values of another set.} 
\CR{Indeed, Eq.\ref{J} is a  map describing the impact of variables in the past onto the future; non-zero elements of $\mathbf{J}_{2,1}$ single out the subset of variables of  $\bo{X}_1$ that has direct future impact on a subset of variables of $\bo{X}_2$.
With LoGo we can identify the channels that  spread instability through the network and we can quantify their effects.}

\CR{Risk and systemic vulnerability are described instead by the  structure of $\mathbf{J}_{2,2}$. } 
Indeed, the expected conditional fluctuations of the variables $\bo{X}_2$  are quantified by the conditional covariance:
\begin{equation}
\mathbf{Cov}\left(\bo{X_2} | \bo{X_1}\right)= \mathbf{J}_{2,2}^{-1} \;\;,
\label{ConditionalCov}
\end{equation}
which involves the term $\mathbf{J}_{2,2}$ only, which therefore describes  propagation of uncertanty across variables.
\CR{Through this term we can link Information Filtering Network with causality relations, indeed Granger causality and transfer entropy are both associated to the ratio of the determinants of two conditional covariances between past and future variables \cite{Granger,Barnett}.
This introduces a novel way to associate causal directionality to  Information Filtering Networks.
}

\subsection{Financial applications: Stress Testing and Risk Allocation}

\subsubsection{Financial applications: Stress Testing}
A typical stress test for financial institutions, required by regulatory bodies, consists in forecasting the effect of severe financial and economic shocks on the balance sheet of a financial institution. 
In this context let's reformulate the previous results by considering $\bo{X}_1$ the set of economic and financial variables that can be shocked and $\bo{X}_2$ the set of the securities held in an institution's portfolio. 
Assuming that  all the changes in the economic and financial variables and in the assets of the portfolio can be modelled as a GMRF, then Eq.\ref{J} represents the distribution of the returns of the portfolio ($\bo{X}_2$) conditional on the realization of the economic and financial shocks ($\bo{X}_1$). 
An approach along similar lines was proposed in \cite{rebonato2010coherent,denevpgm}.
We note that  with the LoGo approach we have a sparse relationship between the financial variables and the securities. This makes calibration more robust and it can be insightful to identify mechanisms of impact and vulnerabilities.  

\subsubsection{Risk Allocation}
A second application is the calculation of conditional statistics in the presence of linear constraints (see \cite{rueheld2005}). In this case we indicate with $\bo{X}$ a set of $p$ random variables associated with the returns in a portfolio of $p$ assets and with $\mathbf{J}$ the associated sparse inverse covariance matrix. Let $\mathbf{w}\in \mathbb{R}^{p\times 1}$ be the vector of holdings of the portfolio, then $\mathbf{w}^\mathsf{T}  \cdot \bo{X}$ is the return of the portfolio. 
An important question in portfolio management is to allocate profits and losses to different assets conditional on a given level of profit or loss, which is equivalent to knowing the distribution of returns conditional on a given level of loss $\bo{X} | \mathbf{w}^\mathsf{T}  \cdot \bo{X} = \bo{L}$. 
More generally we want to estimate $\bo{X} | \bo{A} \cdot \bo{X} = \bo{z}$ where $\bo{A} \in \mathbb{R}^{k\times p}$ is generally a low rank $k$ ($k=1$ in our example) matrix that specifies $k$ hard linear constraints. 
Using the Lagrange Multipliers method (see \cite{strang1986introduction} for an introduction) the conditional mean is calculated as (\cite{rueheld2005}):
\begin{equation} \label{eq:condmean}
\mathbb{E}\left(\bo{X} | \bo{A} \cdot \bo{X} = \bo{z}\right) =  \bo{A}  \mathbf{J}^{-1} \left(\bo{A}  \mathbf{J}^{-1}\bo{A}^\mathsf{T}
\right)^{-1} \bo{z}
\end{equation} 
and the conditional covariance is:
\begin{equation} \label{eq:condvariance}
\mathbf{Cov}\left(\bo{X} | \bo{A} \cdot \bo{X} = \bo{z}\right) = \mathbf{J}^{-1} - \mathbf{J}^{-1} \bo{A}^\mathsf{T} \left(\bo{A}  \mathbf{J}^{-1}\bo{A}^\mathsf{T} \right)^{-1} \bo{A} \mathbf{J}^{-1}\;\;.
\end{equation} 

In case $\mathbf{J}$ is estimated using decomposable  Information Filtering Networks (MST or TMFG) then it can be written as a sum of smaller matrices (as in algorithm \ref{alg:4cliqueTree}) involving cliques and separators:

\begin{equation} \label{eq:J_decomposition}
\mathbf{J} = \sum_{\mathcal{C} \in Cliques} \mathbf{J_\mathcal{C}} - \sum_{\mathcal{S} \in Separators} \mathbf{J_\mathcal{S}}
\end{equation}
This decomposition allows for a sparse and potentially parallel evaluation of the matrix products in Eqs.~\ref{eq:condmean} and \ref{eq:condvariance}.

This framework can therefore be used to build the Profit/Loss (P/L) distribution of a portfolio, conditionally on a number of explanatory variables, and to allocate the P/L to the different assets conditional on the realization of a given level of profit and loss. The solution is analytical and therefore extremely quick. Besides, given the decomposability of the portfolio, Eq.\ref{eq:J_decomposition} allows to calculate important statistics in parallel, by applying the calculations locally to the cliques and to the separators. For instance, it is a simple exercise to show that the unconditional expected P/L and the unconditional volatility can be calculated in parallel by adding the contributions of the cliques and subtracting the contributions of the separators. In summary LoGo provides the possibility to build a basic risk management framework that allows risk aggregation, allocation, stress testing and scenario analysis in a multivariate Gaussian framework in a quick and potentially parallel fashion.

\section{Conclusions}
\label{sec:conclusions}
 We have introduced a methodology, LoGo, that makes use of  Information Filtering Networks to \CR{produce  probabilistic models which are sparse, are associated with high likelihood and are computationally fast making possible their use with very large datasets.}
 \CR{It has been established that  Information Filtering Networks produce sparse structures that well reflect the properties of the underlying complex system \cite{tumminelloetal2005}; however, so far,  Information Filtering Networks has been only used for descriptive purposes, now LoGo provides us an instrument to build predictive models from  Information Filtering Networks opening an entirely new range of perspectives that we have just started exploring. }
 
LoGo produces high-dimensional sparse inverse covariances by using low-dimensional local inversions only, making the procedure computationally very efficient and little sensitive to the curse of dimensionality.  
The construction through a sum of local inversion, which is at the basis of LoGo, makes this method very suitable for parallel computing and dynamical adaptation by local, partial updating. 
We discussed the wide potential applicability of LoGo for sparse inverse covariance estimation, for sparse forecasting models and for financial applications such as stress-testing and risk allocation.

By comparing the results \CR{of LoGo modelling, for financial data,} with a state-of-the-art Glasso procedure we \CR{demonstrated that LoGo can obtain}, in a fraction of the computation time, equivalent or better results with a sparser network structure.
\CR{
However, we observed that when applied to factor models with more than a few factors, LoGo is underperforming with respect to less sparse or dense models. This is probably the consequence of the present LoGo implementations that use  Information Filtering Networks (MST and TMFG) with constrained sparsity (respectively $p-1$ and $3p-6$ edges).
Such a limitation can be easily overcome by constructing more complex networks with larger maximum cliques and larger number of edges. 
A natural extension beyond MST and TMFG would be to use, instead of the present greedy local likelihood maximization under topological constraints, an information criteria (such as Akaike information criterion \cite{akaike1974new,akaike1998information})  which let a chordal graph to be constructed through local moves without constraining a priori its final topological properties. 
This would produce clique forests which generalize the MST and TMFG studied in this paper.
Further extensions could be along the lines of the biologically motivated work \cite{Molinelli13} where ensemble of inference network were explored.
These extensions could increase model robustness; this however would be unavoidably at expenses of computational efficiency. 
The trade-off between efficiency and performance  for the choice of  Information Filtering Networks for LoGo will be the topic of future investigations.
}

The model introduced in this paper is a second-order solution of the maximum entropy problem, resulting in linear, normal multivariate modelling. 
It is however well known that many real systems follow non-linear probability distributions.
Linearity is a severe limitation which can however be overcome in several ways.
For instance, LoGo can be extended to a much broader class of non-linear models  by using the so-called kernel trick \cite{shawe2004kernel}.
Other generalisations to non-linear transelliptical models \cite{liu2012transelliptical} can also be implemented.
\CR{Furthermore, the probability decomposition at the basis of LoGo (Eq.\ref{eq:Factorizing}) is general and can be even applied to non-parametric, non-linear models.}
These would be however, the topics of future works.

 \section*{Acknowledgements}
 T.A. acknowledges support of the UK Economic and Social Research Council (ESRC) in funding the Systemic Risk Centre (ES/K002309/1). 
 TDM wishes to thank the COST Action TD1210 for partially supporting this work. We acknowledge Bloomberg for providing the data.

\appendix


\section{Comparison between LoGo and state-of-the-art sparse Glasso model for non stationary financial data}\label{Sii}
\begin{figure}[t]
\centering
\includegraphics[width=0.7\textwidth]{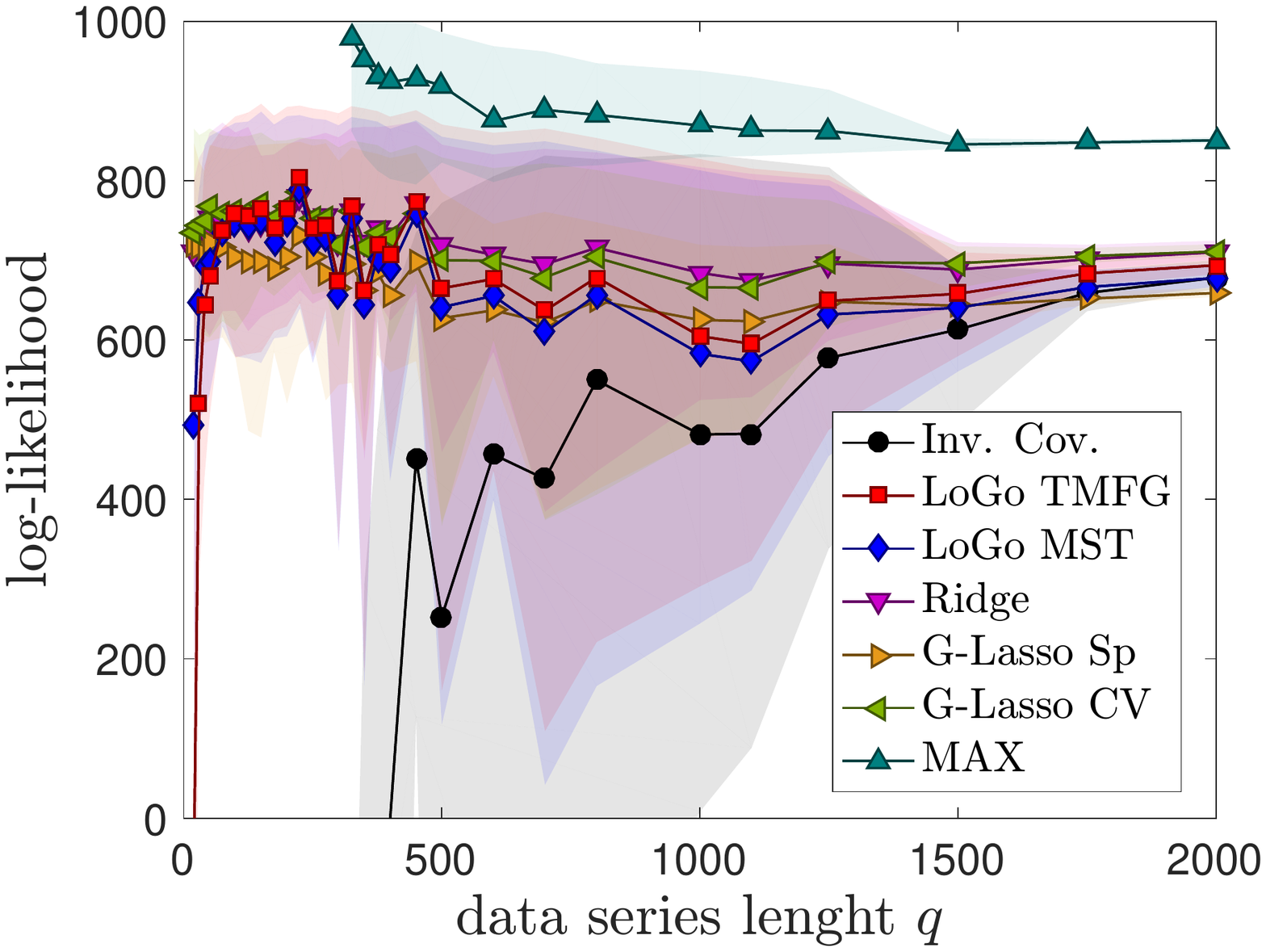}
\caption{
{\bf Demonstration that also for non-stationary financial data LoGo sparse inverse covariance represents the dependency structure better than the complete inverse covariance.}
The figure reports comparisons between log-likelihood for models constructed by using sparse inverse LoGo-TMFG, LoGo-MST and the complete inverse  of the empirical covariance matrix (Inv. Cov.). 
The measures are on $p=300$ off-sample test data-series of different lengths $q$ varying from 20 to 2000.
Inverse matrices are computed on training datasets of the same length.
Data are log-returns sampled from 342 stocks prices of equities traded on the US market during the period 1997-2012.
Symbols correspond to averages over 100 samples generated by picking at random 300 series over the 342 and assigning training and testing sets by choosing at random two consecutive non-overlapping time-windows of length $q$, the shaded bands are the 95\% quantiles.
Testing set is the time-window preceding the training set.
The line on the top, labelled with `MAX', is the theoretical maximum which is the  log-likelihood obtained from the inverse covariance of the testing set. 
}
\label{fig:LL_comparisonWithInverseREAL}\label{fig.S2}
\end{figure}

We investigated the comparison  between LoGo and state-of-the-art sparse Glasso model from  \cite{hsieh2014quic} for the same financial data used for Figs.\ref{fig:LL_comparisonWithGlasso} and Tab.\ref{tab:RealDataShuff} but using the real temporal sequence of the financial data. 
These sequences are non-stationary having varying statistical properties across the time windows. This unavoidably must affect the capability of the model to describe statistically test data from the study of the training data  being the two associated with different temporal states where different events affect the market dynamics. 
Results for the log-likelihood are reported in Fig.\ref{fig:LL_comparisonWithInverseREAL}, these are the same results reported  in Tab.\ref{tab:RealData}.
By comparing with Fig.\ref{fig:LL_comparisonWithGlasso} we observe a much greater overall noise with an interesting collapse of  performances with similar values for all models.
We also observe larger log-likelihoods for shorter time-window. This fact is commented in Section \ref{sec:results}.
Let us notice that these results are in agreement with the finding for the stationary case with LoGo  still performing better or comparably well with respect to Glasso-CV. Computational times and sparsity value are very similar to what reported for the stationary case.

\section{Comparison between LoGo and state-of-the-art sparse Glasso model for sparse factor models}\label{Siii}
\begin{figure}[t]
\centering
\includegraphics[width=0.7\textwidth]{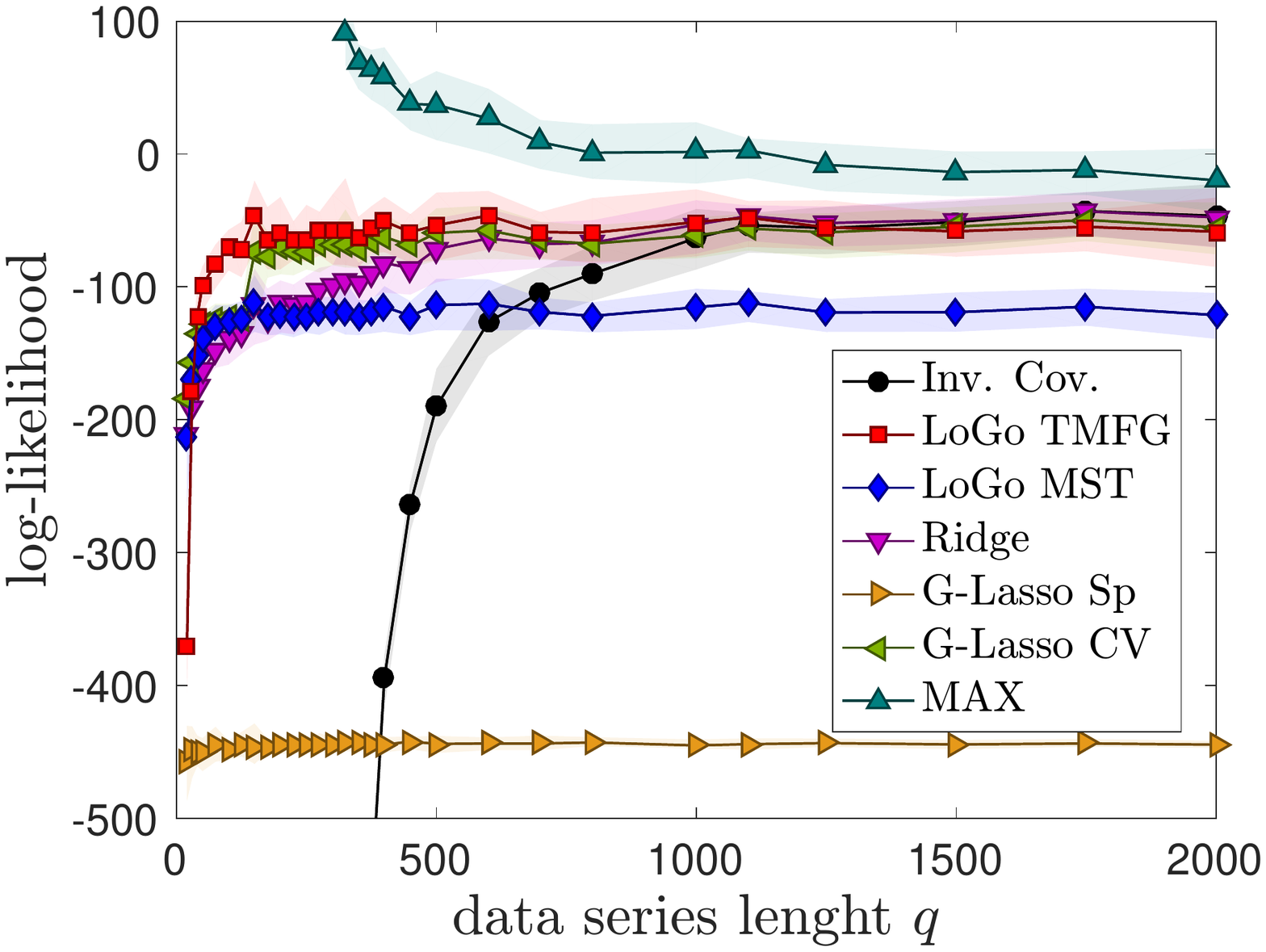}\\
\includegraphics[width=0.7\textwidth]{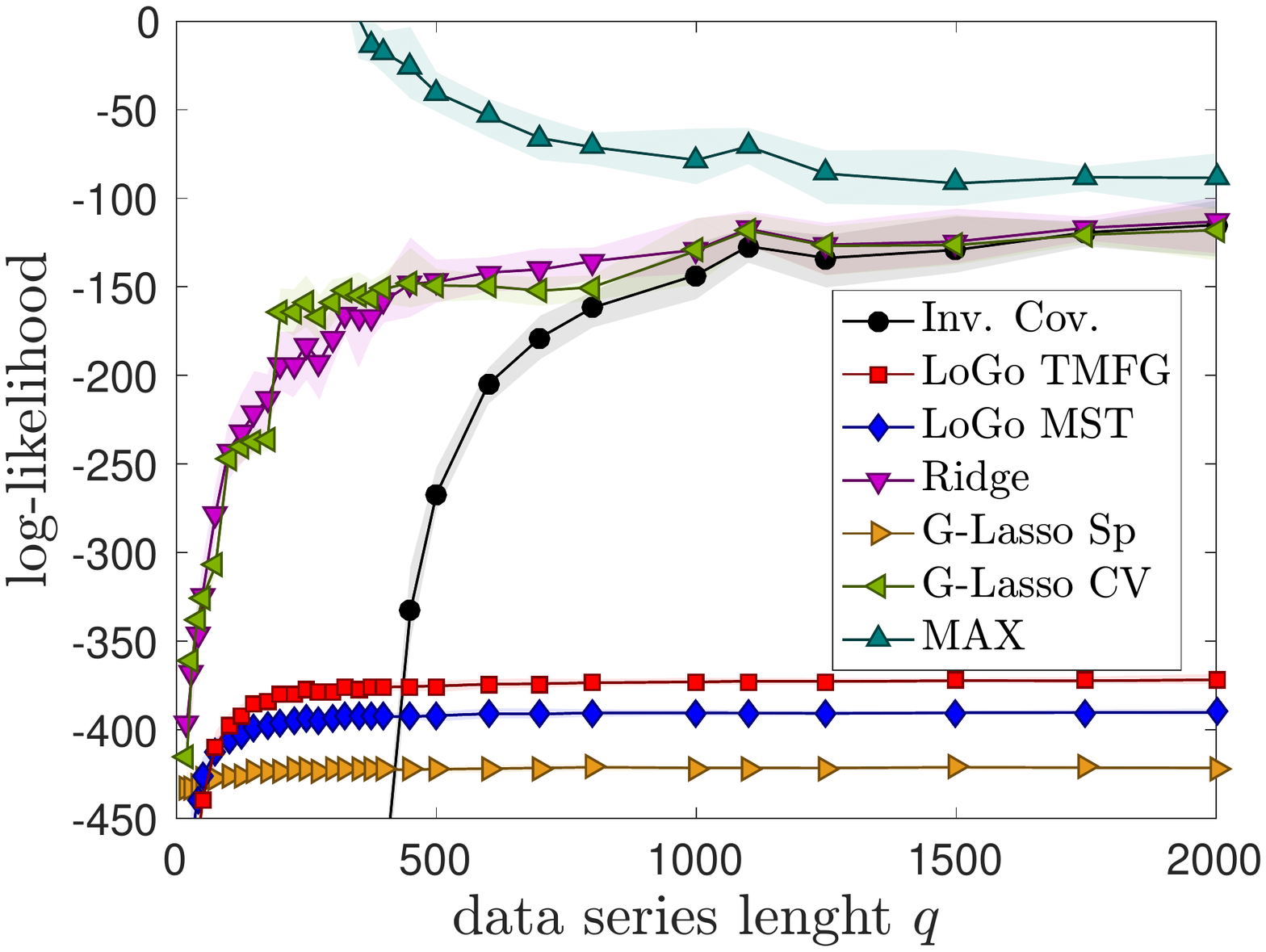}
\caption{
{\bf Demonstration that also for sparse factor models LoGo sparse inverse covariance can perform well when the number of factors is low but underperforms for larger number of factors.}
The figure reports comparisons between log-likelihood for models constructed by using sparse inverse LoGo-TMFG, LoGo-MST and the complete inverse  of the empirical covariance matrix (Inv. Cov.). 
The measures are on $p=300$ off-sample test data-series of different lengths $q$ varying from 20 to 2000.
Inverse matrices are computed on training datasets of the same length.
Plots refer to sparse factor models simulations with 3 (top) and 30 (bottom) factors respectively.
Symbols correspond to averages over 100 samples generated by picking at random 300 series over the 342 and assigning training and testing sets by choosing at random two consecutive non-overlapping time-windows of length $q$, the shaded bands are the 95\% quantiles.
Testing set is the time-window preceding the training set.
The line on the top, labelled with `MAX', is the theoretical maximum which is the  log-likelihood obtained from the inverse covariance of the testing set. 
}
\label{fig:LL_comparisonWithInverseFACTOR}\label{fig.S3}
\end{figure}

Plots for the log-likelihood vs. the time series length for factor models are reported in Fig.\ref{fig:LL_comparisonWithInverseFACTOR} which correspond to the results reported in the paper in Tab.\ref{tab:RealData}. 
We observe that LoGo still performs well and relatively similar to the real data performances when the number of factors is small. 
Conversely, when the number of factors increases LoGo underperforms even the inverse correlation for $q>400$.
Let us note that LoGo is still over performing the Glasso with the same number of parameters.


\end{document}